\documentclass[aps,pra,twocolumn,groupedaddress]{revtex4-2}
\usepackage{
		graphicx,	
		amsmath, breqn, feynmf, lipsum, subcaption}	
		
\newcommand{\UUTH}{Institute for Theoretical Physics and Center for Extreme Matter and Emergent Phenomena, Utrecht University}

\newcommand{\UUEXP}{Debye Institute for Nanomaterials and Center for Extreme Matter and Emergent Phenomena, Utrecht University}

\begin{document}

\title{Dynamics of spontaneous symmetry breaking in a space-time crystal}

\author{J.N. Stehouwer}
\affiliation{\UUTH}

\author{J. Smits}
\affiliation{\UUEXP}

\author{P. van der Straten}
\affiliation{\UUEXP}

\author{H.T.C. Stoof}
\affiliation{\UUTH}

\date{\today}

\begin{abstract}
We present the theory of spontaneous symmetry breaking (SSB) of discrete time translations as recently realized in the space-time crystals of an atomic Bose-Einstein condensate. The non-equilibrium physics related to such a driven-dissipative system is discussed in both the Langevin as well as the Fokker-Planck formulation. We consider a semi-classical and a fully quantum approach, depending on the dissipation being either frequency independent or linearly dependent on frequency, respectively. For both cases, the Langevin equation and Fokker-Planck equation are derived, and the resulting equilibrium distribution is studied. We also study the time evolution of the space-time crystal and focus in particular on its formation and the associated dynamics of the spontaneous breaking of a $Z_2$ symmetry out of the symmetry unbroken phase, i.e., the equilibrium Bose-Einstein condensate before the periodic drive is turned on. Finally, we compare our results with experiments and conclude that our theory provides a solid foundation for the observations.
\end{abstract}

\maketitle

\begin{fmffile}{Diagram}


\section{Introduction} \label{intro}
Statistical physics \cite{stat1, stat2} has been extremely successful at describing both the equilibrium properties as well as the non-equilibrium dynamics of many-body systems. In the latter case this is not only true for small disturbances of a system in equilibrium, that can be treated by linear-response theory and therefore still probe equilibrium fluctuations of the system, but also for far out-of-equilibrium dynamics such as quenches across a phase transition and the ensuing relaxation towards a new equilibrium. At present there is also much interest in studying systems which are constantly out of thermodynamic equilibrium as many examples of this can be found in our mostly classical everyday life, such as in fluids \cite{fluid1}, cells \cite{cells1} and suspensions \cite{colloids1}. 

In addition, the investigation of the non-equilibrium dynamics of quantum many-body systems has obtained a large boost in the last decade from the readily availability of ultra-cold atomic gases in research labs, see for instance Refs.\ \cite{exp1, exp2, exp3, exp4, exp5}. Ultra-cold atomic gases are especially suited for non-equilibrium dynamics as essentially all their characteristics can be manipulated and tuned almost at will. For instance, the magnetic or optical trap in which the atoms are stored can be precisely engineered, allowing even for a control over the dimensionality of the atomic gas that typically affects the dynamics in a dramatic way \cite{1d}. In this manner also a periodic drive can be easily applied such that the system cannot behave as it would do in the standard equilibrium situation as the continuous time-translation symmetry is broken into a discrete symmetry only. Nevertheless a non-equilibrium steady state can occur, in which all the energy gain that is provided by the drive is exactly dissipated away from the system \cite{dissipative}.

An especially interesting example of such driven-dissipative systems are so-called (discrete) time crystals: systems that have spontaneously broken the discrete time-translational symmetry of the periodic drive \cite{wilczek, wilczek2, tc}. Time crystals have now been observed experimentally by multiple groups around the world \cite{zhang, choi, autti, rovny, pal, trger, hemmerich}, but our focus in this paper lies on the experimental realization of a space-time crystal in a Bose-Einstein condensate (BEC) of sodium atoms as described in Refs.\ \cite{expdated, exp}. Notice that only in this case a space-time crystal is formed, which means that on top of the periodic structure in time, with twice the period of the drive, it also possesses a crystalline structure in space. Hence, both the discrete time-translation symmetry as well as the continuous translation symmetry are broken spontaneously at the same time.

In the experiment of interest one starts from a cigar-shaped, and therefore axially symmetric gas, which is evaporatively cooled to such low temperatures that an almost complete Bose-Einstein condensation has occurred. By suddenly changing the radial trap frequency one excites a long-lived radial breathing mode that effectively implements a periodic drive for the axial collective modes of the Bose-Einstein condensate. As a response to this drive exactly one crystalline mode, i.e., a standing wave, is excited in the axial direction. As already mentioned, this mode also has a frequency of precisely half the drive frequency and on the basis of its spontaneously broken symmetries it thus corresponds precisely to a space-time crystal. Most interestingly for our purposes, there exists a (hidden) Ising-like symmetry related to the time evolution of the amplitude of the space-time crystal being roughly speaking either in phase or out of phase with the drive. More precisely, the temporal phase difference of a single realization of the experiment can be either $\phi$ or $\phi + \pi$, where the phase $\phi$ depends on the microscopic details of the experiment and is therefore not universal. This is the spontaneous symmetry breaking of the $Z_2$ symmetry that has been observed recently in Ref.\ \cite{ssb} and that will be investigated theoretically in great detail in this paper.

The layout of the rest of the paper is as follows. In Sec.\ \ref{theory} we explain the theoretical methods that are needed to describe the spontaneous symmetry breaking observed in the experiments. In particular, we will be using a functional formulation of the Schwinger-Keldysh formalism that is very convenient for our purposes. In the end it will lead us to the exact Langevin and Fokker-Planck equations that accurately describe the full quantum dynamics of both the amplitude and the phase of the space-time crystal, including the nonlinear dissipation from the small thermal cloud in the atomic gas. In Sec.\ \ref{twoapproaches}, we then split our discussion into two parts, depending on the assumptions we make about the frequency dependence of the dissipation. We consider two cases that are both allowed by the generalized fluctuation-dissipation theorem that we derived in Sec.~\ref{theory}, namely frequency-independent dissipation and a dissipation rate that depends linearly on the frequency. More physically, the validity of these two cases depends on the time scale of interest in the experiment, describing either prethermalization or thermalization, respectively. We investigate the Langevin and Fokker-Planck equations in each of the two approximations. In particular, we will look at the equilibrium solutions, but also discuss numerical solutions of the Fokker-Planck equation that explicitly show the dynamics of the spontaneous symmetry breaking taking place in the Bose-Einstein condensate after the drive is turned on and the space-time crystal is formed out of quantum and thermal fluctuations. After that, we are in a position to also compare our results with experiments in Sec.\ \ref{experiments}. In Sec.\ \ref{conclusion} we end the paper with conclusions and an outlook.

\section{Schwinger-Keldysh set-up} \label{theory}

The starting point for this paper comes from the results of Ref.\ \cite{exp}. In that paper, the average dynamics of the space-time crystals were found to be described very accurately by the Hamiltonian
\begin{equation} \label{Hamiltonian}
    \hat{H} = -\hbar \delta \hat{a}^\dagger \hat{a} + \frac{\hbar \omega_D A_D}{8} (\hat{a}^\dagger \hat{a}^\dagger + \hat{a} \hat{a}) + \frac{\hbar g}{2} \hat{a}^\dagger\hat{a}^\dagger \hat{a} \hat{a},
\end{equation}
where $\delta$ is the detuning from resonance, $\omega_D$ is the driving frequency, $A_D$ is the relative driving amplitude and $\hat{a}^{(\dagger)}$ is the annihilation (creation) operator of a quantum in the resonant axial mode. Note that the detuning enters this Hamiltonian, as it describes the slow dynamics in the frame rotating along with the drive. It is in this rotating frame that the Hamiltonian becomes in a good approximation time independent and that prethermalization can therefore occur. Furthermore, there is the complex interaction parameter $g = g' + i g'' = |g| e^{i\phi_g}$, which was introduced phenomenologically to counter the exponential growth of the space-time crystal due to the presence of the drive and to allow for a relaxation into a long-lived prethermal state as observed experimentally. Phenomenologically, this nonlinear interaction term leads to an excellent agreement with the (average) growth and relaxation dynamics observed experimentally. In particular, the nonlinear dissipation due to $g''<0$ is required, and the more common linear dissipation does not work in this case. Since the Hamiltonian has an imaginary part and is thus non-hermitian, we expect that also noise will arise because of the fluctuation-dissipation theorem. Therefore, we will expand and support the model of Ref.\ \cite{exp} by microscopically deriving this imaginary piece and its associated noise. We will do this via a non-linear coupling with a heat bath, resulting in both the imaginary part of the Hamiltonian, as well as the noise that satisfies a generalized fluctuation-dissipation theorem.

An in our opinion elegant way to develop our theory is by using the functional techniques of the Schwinger-Keldysh formalism. With that, we will derive an effective action that precisely corresponds to the Hamiltonian of Eq.\ (\ref{Hamiltonian}) and that automatically also incorporates the fluctuations generated by the non-hermitian term in the Hamiltonian. As the dynamics of the drive is fully coherent, we focus for simplicity first on the derivation of the nonlinear interaction term by considering the case of $A_D=0$ and $\omega_D=0$, i.e., without the drive. After that has been achieved, the effect of the drive can still be easily incorporated exactly into the resulting Langevin and Fokker-Planck equations. We hope that this makes the following discussion more clear by avoiding unnecessary long expressions.

Let us first derive the microscopic action corresponding to the non-linear coupling with the heat bath. The probability distribution $P[a^*,a;t]$ for the space-time crystal to be in the coherent state $|a \rangle$ of the annihilation operator $\hat{a}$ at time $t$ can be expressed in terms of the matrix elements $\langle a; t | a_0; t_0 \rangle $ and its complex conjugate $\langle a; t | a_0; t_0 \rangle^* = \langle a_0; t_0 | a; t \rangle $, where $| a; t \rangle$ and $| a_0; t_0 \rangle$ are the coherent states at the present time $t$ and at the initial time $t_0$, respectively. For these matrix elements, the `path'-integral expressions going through all possible field configurations are calculated, e.g.\ in Refs.\ \cite{stoof, negele}. Because of the fact that the complex conjugate has an opposite time ordering, the actions corresponding to these elements will have opposite evolutions in time. This leads to the introduction of the Schwinger-Keldysh contour $\mathcal{C}$, which obeys $\int_{\mathcal{C}} dt' = \int_{t_0}^{t} dt_+ + \int_{t}^{t_0} dt_-$. For our purposes of calculation equal-time correlation functions, we are allowed to deform the contour such that it encircles the complete time axis by taking the limits $t_0 \to - \infty$ and $t \to \infty$. For more details, we refer to Ref.\ \cite{toymodel}.

To reproduce the non-hermitian interaction term in the Hamiltonian we first of all make use of the fact that for ultra-cold atoms a possible frequency dependence of the interaction comes mostly from the dependence of the interatomic collisions on the center-of-mass energy. Therefore, we imagine a decoupling of the interaction in the particle-particle channel and consider a linear interaction of the heat bath with $\psi(t)\psi(t)$, where $\psi(t)$ refers to the field of the space-time crystal. This implies that two quanta from the resonant axial mode can combine into a single quantum of the heat bath and, because of time-reversal symmetry of the microscopic action, that one quantum of the heat bath can also decay into two quanta of the resonant axial mode. Since the quadratic piece of the Hamiltonian is not affected by these scattering processes with the heat bath, we can in more detail write for the microscopic action that is the starting point of our discussion
\begin{widetext}
\begin{equation}
\begin{aligned}
    S[\psi^*,\psi, \psi^*_R, \psi_R] &= - \frac{1}{\sqrt{V}} \sum_{\mathbf{k}} \int_{\mathcal{C}} dt \bigg \{ t(\mathbf{k}) \psi(t) \psi(t) \psi^*_{\mathbf{k}}(t) + t^*(\mathbf{k}) \psi_{\mathbf{k}}(t) \psi^* (t) \psi^*(t) \bigg \} \\
    & + \int_{\mathcal{C}} dt \psi^*(t) \left(i \hbar \frac{\partial}{\partial t} + \hbar \delta \right) \psi(t) + \sum_{\mathbf{k}} \int_{\mathcal{C}} dt \psi^*_{\mathbf{k}}(t) \left(i\hbar \frac{\partial}{\partial t} - \epsilon (\mathbf{k}) \right) \psi_{\mathbf{k}}(t),
\end{aligned}
\end{equation}
\end{widetext} 
where $\psi_R(\mathbf{x},t) = \sum_{\mathbf{k}} \psi_{\mathbf{k}} e^{i\mathbf{k} \cdot \mathbf{x}} / \sqrt{V}$ is the reservoir field describing the homogeneous heat bath with volume $V$, $t({\bf k})$ are the `tunneling' matrix elements or scattering amplitudes between the space-time crystal and the heat bath, $\epsilon({\bf k})$ denotes the energy of a reservoir state with wave vector ${\bf k}$, and $\mathcal{C}$ is the deformed Schwinger-Keldysh contour that was introduced above. Note that since we are dealing in Eq.~(1) with a single-mode Hamiltonian for the space-time crystal, there is no position dependence in the field $\psi(t)$ in contrast to the reservoir field $\psi_R({\bf x},t)$, which in the thermodynamic limit $V \rightarrow \infty$ describes the continuum required for the occurrence of dissipation and the resulting `arrow of time'. 

\begin{figure}[b]
\setlength{\unitlength}{2mm}\large
\begin{fmfgraph*}(30,30) \fmfpen{thick}
\fmfbottom{i1,i2} \fmftop{o1,o2}
\fmf{fermion}{i1,v1} \fmf{fermion}{i2,v1}\fmf{fermion}{v2,o1} \fmf{fermion}{v2,o2}
\fmf{photon}{v1,v2} \fmfdot{v1,v2}
\fmflabel{$t'$}{v1}
\fmflabel{$t$}{v2}
\end{fmfgraph*}
\caption{Feynman diagram of the effective interaction in the space-time crystal. The wiggly line corresponds to the Green's function or propagator of the reservoir.}
\label{fig:greensfunction}
\end{figure}
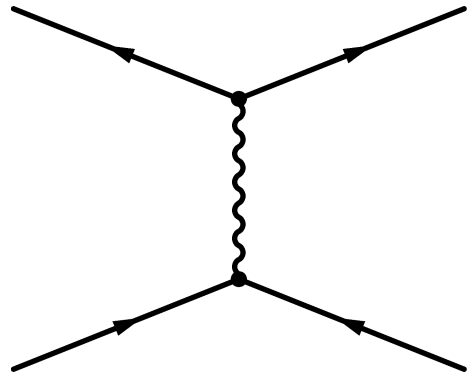

The next step is to integrate out the heat bath since the action is quadratic in $\psi_{\mathbf{k}}$. This gives the desired Schwinger-Keldysh action for the full quantum dynamics of the space-time crystal as
\begin{eqnarray}
    S[\psi^*,\psi] &=& \int_{\mathcal{C}} dt \psi^*(t) \left(i \hbar \frac{\partial}{\partial t} + \hbar \delta \right) \psi(t) \\
    &-& \int_{\mathcal{C}} dt \int_{\mathcal{C}} dt' \psi^*(t) \psi^*(t) \frac{\hbar g(t,t')}{2} \psi(t')  \psi(t'). \nonumber
\end{eqnarray}
Here we introduced the very important two-point function on the Schwinger-Keldysh contour $\hbar g(t,t') = (2/\hbar V) \sum_{\mathbf{k}} t^*(\mathbf{k}) G(\mathbf{k}; t,t') t(\mathbf{k})$, which determines the coupling strength of the effective interaction. Furthermore, the reservoir Green's function $G(\mathbf{k}; t, t')$ satisfies $( i\hbar \partial / \partial t - \epsilon (\mathbf{k})) G(\mathbf{k}; t, t') = \hbar \delta(t,t')$. A Feynman diagram corresponding to the effective interaction is shown in Fig.\ \ref{fig:greensfunction}. Note in particular the appearance of a fourth-order term in $\psi$ in our effective action, which would become a quadratic selfenergy term \cite{toymodel} if we would have used a linear coupling with the heat bath as is common in the Caldeira-Leggett-like models of macroscopic quantum mechanics \cite{tony}.

We have almost completed the derivation of the effective non-equilibrium action. The last step is projecting the two (forward and backward) branches of the Schwinger-Keldysh contour onto the real axis to obtain the real-time dynamics. This is achieved by making the transformation $\psi(t_{\pm}) = a(t) \pm \xi(t)/2$, where $a(t)$ represent the average or `classical' part of the $\psi$ field and $\xi(t)$ the fluctuations, as we will see explicitly in a moment. Because of the fourth-order nature of the effective interaction, we can in principle get fluctuations up to fourth order in $\xi$, instead of only up to quadratic order as in the Caldeira-Leggett-like models with a linear coupling to the heat bath \cite{toymodel}. Nonetheless, we can still work out the action. Up to second order in the fluctuations, we find
\begin{widetext}
\begin{equation} \label{action}
\begin{aligned}
    S^{\rm eff}[a^*,a; \xi^*,\xi] &=  \int dt \bigg \{ a^*(t) \bigg ( i \hbar \frac{\partial}{\partial t} + \hbar \delta \bigg ) \xi(t) + \xi^*(t) \bigg ( i\hbar \frac{\partial}{\partial t} + \hbar \delta \bigg ) a(t) \bigg \}\\
    & - \int dt \int dt' \bigg \{ a^*(t) a^*(t)\hbar g^{(-)}(t-t') a(t')\xi(t') + \xi^*(t) a^*(t) \hbar g^{(+)}(t-t') a(t')a(t')   \bigg \} \\
    & - \int dt \int dt' \xi^*(t) a^*(t) \hbar g^K (t-t') a(t') \xi(t'),
\end{aligned}
\end{equation}
\end{widetext}
where the retarded, advanced, and Keldysh functions, $g^{(\pm)}(t-t')$ and $g^{K}(t-t')$ respectively, are related to the analytic (known as bigger and lesser) pieces of $g(t,t')$ via
\begin{dmath}
    g^{(\pm)}(t-t') = \pm \theta(\pm(t-t')) \left( g^>(t-t') - g^<(t-t') \right),
\end{dmath}
and
\begin{dmath}
    g^{K}(t-t') = g^>(t-t') + g^<(t-t'),
\end{dmath}
with $\theta(t)$ the usual Heaviside step function. Using these definitions and our explicit expression for the two-point function on the Schwinger-Keldysh contour $g(t,t')$ we can now prove that the Fourier transforms of the above retarded, advanced, and Keldysh functions are related by the famous fluctuation-dissipation theorem 
\begin{dmath} \label{fltheorem}
    g^K(\omega) = \left( g^{(+)}(\omega) - g^{(-)}(\omega) \right)
    (1 + 2N(\omega)),
\end{dmath}
where we introduced the Bose or Planck distribution $N(\omega)=1/(e^{\hbar\omega/k_B T}-1)$ and the temperature $T$ of the heat bath. This crucial relation will ultimately ensure that our probability distribution $P[a^*,a;t]$ relaxes to the correct physical equilibrium.

The effective action also has a third-order contribution, namely 
\begin{eqnarray}
-\frac{1}{8} \int dt \int dt' & \hspace{-0.5cm} \bigg ( \xi^*(t) \xi^*(t) \hbar g^{(-)}(t-t') a(t')\xi(t') \nonumber \\ &+ \xi^*(t) a^*(t) \hbar g^{(+)}(t-t') \xi(t')\xi(t')   \bigg ) \nonumber, 
\end{eqnarray}
whereas the fourth-order term turns out to be exactly zero. In the remainder of this paper, we will ignore these third-order terms in the fluctuations, because they lead to quantum corrections to the equations of motion resulting from the Hamiltonian in Eq.~(1), that we already know to be  unimportant for an understanding of the experiments of interest to us here. In the literature the neglect of these higher-order fluctuations is sometimes also known as the truncated Wigner approximation \cite{wig1, wig2}.

There are now two ways to proceed and extract the non-equilibrium dynamics from the effective action in Eq.~(\ref{action}). First, we can integrate out the fluctuations and arrive at an effective action for the field $a(t)$ alone. Quantizing this action and writing down the associated Schr\"odinger equation leads then to the Fokker-Planck equation for the probability distribution $P[a^*,a;t]$. Second, we can first derive the Langevin equation for $a(t)$ and then obtain the desired Fokker-Planck equation from that. We will follow the latter approach here, because it avoids some operator-ordering problems that arise in the former approach. But for consistency reasons and as a check on our calculations, we illustrate also the direct derivation of the Fokker-Planck equation in the following section. 

To obtain the Langevin equation we decouple the quadratic term in the fluctuations by a Hubbard-Stratonovich transformation, see e.g.\ Ref.\ \cite{stoof}. In this case this can most conveniently be achieved by multiplying the integrand of our functional integral over $a(t)$ and $\xi(t)$  by the Gaussian integral
\begin{equation}
    1 = \int d[\eta^*] d[\eta] \exp \left( \frac{i S^{\rm eff}[\eta^*,\eta]} {\hbar} \right),
\end{equation}
where we take
\begin{widetext}
\begin{equation}
\begin{aligned}
S^{\rm eff}[\eta^*,\eta] = \int dt \int dt' \bigg ( \eta^*(t) -\int dt'' \xi^*(t'')a^*(t'') \hbar g^K(t''-t) \bigg ) \frac{{g^K}^{-1}(t-t')}{\hbar} \bigg ( \eta(t') - \int dt'' \hbar g^K (t'-t'') \xi(t'')a(t'') \bigg ).
\end{aligned}
\end{equation}
This will by construction precisely cancel out the quadratic term in $\xi$ of our effective action, and we thus end up with an action that is completely linear in $\xi$. Adding the above $S^{\rm eff}[\eta^*,\eta]$ to our effective action in Eq.\ (\ref{action}) indeed leads to
\begin{equation} \label{actioneff}
\begin{aligned}
S^{\rm eff}[a^*, a, \xi^*, \xi, \eta^*, \eta] & = 
\int dt \bigg \{ a^*(t)
\left( i\hbar \frac{\partial}{\partial t} + \hbar \delta \right) - \int dt' a^*(t') a^*(t') \hbar g^{(-)}(t'-t) a(t) - a(t) \eta^* (t) \bigg \} \xi(t) \\
& + \int dt \xi^*(t) \bigg \{  \left( i\hbar \frac{\partial}{\partial t} + \hbar \delta \right) a(t) - \int dt' a^*(t) \hbar g^{(+)}(t-t') a(t') a(t')
- \eta (t) a^*(t) \bigg \}  \\
& + \int dt \int dt' \eta^*(t) \frac{{g^K}^{-1}(t-t')}{\hbar} \eta(t').  
\end{aligned}
\end{equation}
\end{widetext}
Because the effective action has become linear in the fluctuations after the Hubbard-Stratonovich transformation, integrating out $\xi(t)$ now leads to $\delta$ functionals in a similar manner as in the well-known identity $\int dk e^{ikx} = 2\pi \delta(x)$. As a result we can immediately read off from Eq.~(\ref{actioneff}) that our Langevin equation becomes
\begin{eqnarray} \label{langevin}
    i\hbar \frac{\partial}{\partial t} a(t) &=& -\hbar \delta a(t) + \int dt' a^*(t) \hbar g^{(+)}(t-t') a(t') a(t') \nonumber \\
    &+& \eta(t) a^*(t) ,
\end{eqnarray}
together with the complex conjugate equation for $a^*(t)$. Furthermore, we also conclude from our effective action in Eq.~(\ref{actioneff}) that the noise correlations obey
\begin{align}
  & \langle \eta (t) \eta(t') \rangle = \langle \eta^* (t) \eta^*(t') \rangle = 0 , \\
  & \langle \eta^* (t) \eta(t') \rangle = i\hbar^2 g^K(t-t'). \label{corrfun}
\end{align}
Note in particular that the Langevin equation contains multiplicative noise, due to the fact that the dissipation occurs in the non-linear term in our case. As always in physics, this multiplicative noise must be interpreted as Stratonovich noise that leads also to noise-induced drift terms as we show next. 

\section{Frequency dependence of the dissipation} \label{twoapproaches}

At this point, we will split our discussion up into two parts in order to further analyze the general Langevin equations given by Eq. (\ref{langevin}). We will do this by calculating the equations of motion for the averages $\langle a(t) \rangle$ and $\langle a^*(t) \rangle$ and the fluctuations $\langle a^*(t) a(t) \rangle$, and then inferring the Fokker-Planck equation from these equations of motion. We can, as a first approximation, assume that the dissipation doesn't have any frequency dependence, meaning that the interactions act locally in time. Our second method will be the situation where we allow for a linear frequency dependence. These two options will be our two different approaches.

The goal of our first approximation, which is a semi-classical approach, is to describe the relatively short time scales explored by the experiments in which the prethermalization of the space-time crystal occurs. This will already allow us to study the interesting physics associated with the spontaneous symmetry breaking, as we will see explicitly in Sec.~\ref{numerics}, and to make a direct comparison with experiments in Sec.~\ref{experiments}. The second approach builds upon the first one and uses its techniques to describe the full quantum dynamics in the low-frequency limit that is needed to show that on the longest time scales the system ultimately relaxes to the correct thermal equilibrium. These long time scales have not been explored experimentally and are also expected to be very difficult to explore in practice. However, we add this quantum approach here for completeness and to show the consistency of our theory. In particular, it shows the crucial role played by the fluctuation-dissipation theorem that is automatically incorporated in the Schwinger-Keldysh formalism as we have seen. 

\subsection{Semi-classical dynamics} \label{whitenoise}

So let us start with the semi-classical approach. We make the approximation that the dynamics is dominated by a single frequency $\bar{\omega}$ and the fluctuation-dissipation theorem can be evaluated at twice that typical frequency. In the case without a drive, that we are still considering at the moment, we have for instance $\bar{\omega} = -\delta$, where it should be noted that with $\omega_D = 0$ we have that $-\delta$ just equals the Bogoliubov excitation frequency of the resonant axial mode. This means that our fluctuation-dissipation theorem takes the form
\begin{equation}
    g^K = 2ig''(1+2N(2\bar{\omega})) \simeq 2ig'' \frac{k_B T}{\hbar \bar{\omega}}, \label{fl}
\end{equation}
where $g^K \equiv g^K (2\bar{\omega})$, $g = g' \pm ig'' \equiv g^{(\pm)}(2\bar{\omega})$, and for consistency we also applied the semi-classical approximation to the Bose distribution. Stated in the time domain, we thus have $g^{(\pm, K)}(t-t') = g^{(\pm, K)} \delta(t-t')$ and all memory effects have disappeared. Note that this local or Markovian approximation is indeed only valid at shorter time scales as the Bose distribution diverges for $\omega \rightarrow 0$. At the longest time scales a full quantum approach is therefore needed, which we develop in Sec.~\ref{colourednoise}.  

\subsubsection{Langevin formulation} \label{L formulation}

In the semi-classical approximation a general solution of the Langevin equation in Eq. (\ref{langevin}) reads
\begin{dmath}
   a(t) = e^{i\delta t} \bigg \{ a(0) - i g \int_0^t dt'  a^*(t')a(t')a(t')e^{-i\delta t'} - \frac{i}{\hbar} \int_{0}^{t} dt' \eta(t')a^*(t')e^{-i\delta t'} \bigg \},
\end{dmath}
and similarly for $a^*(t)$. With that, we can calculate the equal-time correlation functions that contain $\eta(t)$, $a(t)$, and $a^*(t)$. The results of interest to us become
\begin{align}
& \langle \eta(t) a^*(t) \rangle =  -\frac{\hbar g^K}{2} \langle a(t) \rangle ,  \\
& \langle \eta(t) a^*(t) a^*(t) \rangle =  -\hbar g^K \langle a^*(t) a(t) \rangle ,
\end{align}
where we used that $g^K$ is purely imaginary as explicitly shown in Eq.~(\ref{fl}), and $\langle \eta(t) \rangle = \langle \eta^*(t) \rangle = 0$.
With these results, we conclude that 
\begin{dmath}\label{langevinwhite}
    i\hbar \frac{\partial}{\partial t} \langle a(t) \rangle = -\hbar \delta \langle a(t) \rangle + \hbar g \langle a^*(t) a(t) a(t) \rangle - \frac{\hbar g^K}{2} \langle a(t) \rangle,
\end{dmath}
and the complex conjugate equation for $\langle a^*(t) \rangle$. In addition the fluctuations are determined by
\begin{dmath}\label{eomwhite}
    i\hbar \frac{\partial}{\partial t} \langle a^*(t) a(t) \rangle = 2i\hbar g'' \langle a^*(t) a^*(t) a(t) a(t) \rangle - 2\hbar g^K \langle a^*(t) a(t) \rangle.
\end{dmath}
This is the full set of equations of motion in the semi-classical approach, where the term on the right-hand side of Eq.~(\ref{langevinwhite}) proportional to $g^K$ is the noise-induced drift term from the Stratonovich noise.

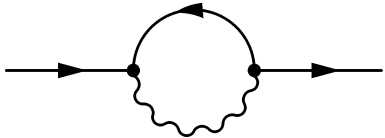
\begin{figure} 
\setlength{\unitlength}{2mm}\large
\begin{fmfgraph*}(25,25)
   \fmfpen{thin}
   \fmfleft{i}
   \fmfright{o}
   \fmf{fermion}{i,v1}
   \fmf{fermion}{v2,o}
   \fmf{fermion, right, tension=0.05}{v2,v1}
    \fmf{photon, right, tension =0.05}{v1,v2}
   \fmfdot{v1,v2}
\end{fmfgraph*}
\caption{Feynman diagram of the scattering process involved in the right-hand side or `collision integral' of the Boltzmann equation. A straight line represents $\langle a^*(t) a(t) \rangle$ and a wiggly line $\langle a^*_k(t) a_k(t) \rangle$.}
\label{fig:feynmandiagram}
\end{figure}

For the physical interpretation of Eq.~(\ref{eomwhite}) we now show that it corresponds to the Boltzmann equation. The occupation numbers $N(t)$ of the space-time crystal are given by \cite{toymodel}
\begin{dmath}
    \langle a^*(t) a(t) \rangle = N(t) + \frac{1}{2} ,
\end{dmath}
containing explicitly both thermal and quantum fluctuations, respectively. Using Wick's theorem \cite{stoof, negele} to work out the four-point correlation function, we are able to write
\begin{dmath} \label{boltzmanneq}
    \frac{\partial}{\partial t} N(t) = - \Gamma \left( N(t) + \frac{1}{2} \right) \bigg \{ \left( N(t)+ \frac{1}{2} \right) - \frac{k_B T}{\hbar\bar{\omega}} \bigg \},
\end{dmath}
where $\Gamma \equiv -4 g''$ is a rate of decay that determines the relaxation of the occupation numbers towards the correct semi-classical equilibrium. It is important to remember that we are still discussing the situation without the presence of a drive, and the equilibrium we have obtained here will therefore later on in Sec.~\ref{numerics} be used as an initial state for the numerical solution of the Fokker-Planck equation that describes most clearly the spontaneous breaking of the $Z_2$ symmetry after the drive is turned on. Before we start discussing this Fokker-Planck treatment, we would like to point out that we can write the above rate equation into a more familiar Boltzmann form as
\begin{dmath}
    \frac{\partial}{\partial t} N(t) = - \Gamma \langle a^*(t) a(t) \rangle \bigg \{ N(t) (1 + N_k) - N_k (1 + N(t)) \bigg \},
\end{dmath}
with $N_k$ describing the equilibrium occupation numbers of the reservoir, the momentum ${\bf k}$ being fixed by the energy conservation $\epsilon({\bf k}) = \hbar \bar{\omega}$. Note in particular the overall factor $\langle a^*(t) a(t) \rangle$ on the right-hand side that is due to the multiplicative noise and effectively renormalizes the rate for quanta of the space-time crystal to scatter into and out of the heat bath. A Feynman diagram of the process involved is shown in Fig.~\ref{fig:feynmandiagram}. 

\end{fmffile}

\subsubsection{Fokker-Planck formulation} \label{FP formulation}

We now turn to the Fokker-Planck description of the obtained Langevin dynamics. We can show that the term $2i\hbar g'' \langle a^* a^* a a \rangle$ on the right-hand side of Eq.~(\ref{eomwhite}) is taken into account by the streaming terms of the Fokker-Planck equation that follow from Eq.~(\ref{langevinwhite}). It therefore does not lead to a contribution in the diffusive part of the Fokker-Planck equation. As a result the diffusion term only has to reproduce the two terms proportional to $g^K$, i.e., the last term in the right-hand side of Eq.~(\ref{langevinwhite}) and of Eq.~(\ref{eomwhite}) that both arise from the multiplicative noise. This completely fixes the diffusion term of the Fokker-Planck equation to be
\begin{eqnarray}
&& - \frac{\hbar g^{K}}{2} \bigg [ \frac{\partial}{\partial a^*} \bigg \{ |a|^2 \frac{\partial}{\partial a} P[a^*,a;t] \bigg\} \nonumber \\
&& \hspace{1.0cm} + \frac{\partial}{\partial a} \bigg \{ |a|^2 \frac{\partial}{\partial a^*} P[a^*,a;t] \bigg \} \bigg ]  . \nonumber
\end{eqnarray}
Specifically for our case, we had to choose a symmetric diffusion operator.
Together with the streaming terms, which follow directly from Eq.~(\ref{langevinwhite}), we find that the Fokker-Planck equation becomes
\begin{eqnarray} \label{fpwhite}
    i \hbar \frac{\partial}{\partial t} P[a^*,a;t] &=& -\frac{\partial}{\partial a}\bigg\{ (-\hbar \delta + \hbar g |a|^2) a P[a^*,a;t] \bigg\} \nonumber \\
    &+& \frac{\partial}{\partial a^*} \bigg\{ (-\hbar \delta + \hbar g^{*} |a|^2) a^* P[a^*,a;t] \bigg\} \nonumber \\
    &-& \frac{\hbar g^{K}}{2} \frac{\partial}{\partial a^*} \bigg \{ |a|^2 \frac{\partial}{\partial a} P[a^*,a;t] \bigg \}  \nonumber \\ 
    &-& \frac{\hbar g^{K}}{2} \frac{\partial}{\partial a} \bigg \{ |a|^2 \frac{\partial}{\partial a^*} P[a^*,a;t] \bigg \} .
\end{eqnarray}
From the Langevin equation and the resulting equations of motion for the average moments $\langle a(t) \rangle$ and $\langle a^*(t) a(t) \rangle$ we have found the Fokker-Planck description that is fully equivalent to the Langevin dynamics. This is the most common procedure in non-equilibrium statistical physics. However, as already mentioned, we can also derive the Fokker-Planck equation directly from the effective action, which is a convenient consistency-check for our theory. This is what we briefly present now.\\

Performing the Gaussian integration over the fluctuation field $\xi(t)$, we obtain from Eq.~(\ref{action}) the desired effective action of the $a(t)$ field alone
\begin{eqnarray} \label{fpdirectaction}
  &&  S^{\rm eff}[a^*,a] \equiv \int dt L(t) \\
    && = \int dt \frac{1}{\hbar g^K |a(t)|^2}
     \bigg \lvert \bigg (i\hbar \frac{\partial}{\partial t} +\hbar \delta -\hbar g |a(t)|^2 \bigg)  a(t) \bigg \rvert ^2 \nonumber.
\end{eqnarray}
Since we still have to perform a functional integral over $a(t)$ to obtain the partition function, this effective action corresponds to the quantum theory of a `classical' system with the Lagrangian $L(t)$. The Schr\"odinger equation of this quantum theory is exactly the Fokker-Planck equation we are interested in. To obtain it we thus need to quantize the theory of Eq.~(\ref{fpdirectaction}). In particular, the canonical momentum $\pi(t)$ of $a(t)$ satisfies
\begin{dmath}
    \pi(t) = \frac{i \hbar}{\hbar g^K |a(t)|^2} \bigg ( -i\hbar \frac{\partial}{\partial t} +\hbar \delta -\hbar g^{*} |a(t)|^2 \bigg ) a^*(t), 
\end{dmath}
and the complex conjugate for the canonical momentum $\pi^*(t)$ of $a^*(t)$. The rest of this derivation goes alone the same way as the discussion of Ref.~\cite{toymodel}, i.e., we can now construct the Hamiltonian $H(t)$ from the Lagrangian $L(t)$ to obtain
\begin{eqnarray} \label{ham}
    H(t) &=&  \frac{\pi(t)}{i\hbar} ( -\hbar \delta + \hbar g |a(t)|^2)a(t) \\
    &-&  \frac{\pi^*(t)}{i\hbar}(-\hbar \delta +\hbar g^{*} |a(t)|^2)a^*(t) +  \frac{g^K |a(t)|^2}{\hbar} \lvert \pi(t)\rvert ^2. \nonumber
\end{eqnarray}
After requiring the usual commutation relations between coordinates and canonical momenta, the Schr\"{o}dinger equation
\begin{dmath}
    i \hbar \frac{\partial} {\partial t} P[a^*,a;t] = H P[a^*,a;t],
\end{dmath}
then indeed exactly reproduces the Fokker-Planck equation of Eq.~(\ref{fpwhite}). As it stands the Hamiltonian in Eq.~(\ref{ham}) has clearly operator-ordering problems but these have fortunately already been resolved from the Langevin derivation that we carried out first. Interestingly, normal ordering is not appropriate for the diffusion term in the Fokker-Planck equation, which is a consequence of the Stratonovich nature of the multiplicative noise that gives rise to a noise-induced drift term. Normal ordering would lead to the more mathematical It\^o interpretation of the multiplicative noise. 

Of course, we can also rederive the Langevin equations from the Fokker-Planck equation. Taking moments of the Fokker-Planck equation and integrating by parts nicely reproduces the desired equations of motion in Eqs.~(\ref{langevinwhite}) and (\ref{eomwhite}). From this we can conclude that the Langevin equation and the Fokker-Planck equation are really two sides of the same coin. As a result  the equilibrium probability distribution of the Fokker-Planck equation in Eq.~(\ref{fpwhite}) should agree with the equilibrium occupation numbers obtained from the Boltzmann equation in Eq.~(\ref{eomwhite}). It can therefore be proven that the equilibrium probability distribution is equal to the semi-classical ideal-gas solution
\begin{equation} \label{gaussdistribution}
    P[a^*,a] = \frac{\hbar \bar{\omega}}{\pi k_B T} \exp \bigg ( - \frac{\hbar \bar{\omega}}{k_BT}  |a|^2 \bigg ),
\end{equation}
a fact that is shown in App.~\ref{appA} in more detail.

All the above was achieved without the drive terms $\hbar \omega_D A_D (\hat{a}^\dagger \hat{a}^\dagger + \hat{a} \hat{a})/8$ in our starting Hamiltonian in Eq.~(\ref{Hamiltonian}). Adding the drive to the system is actually the key ingredient to observe the symmetry breaking we are interested in, since this is the only term that has explicit phase dependence. More precisely, the drive breaks the $U(1)$ phase symmetry down to a $Z_2$ symmetry $\hat{a} \rightarrow - \hat{a}$ only. Because the time evolution due to the drive is unitary it is straightforward to add these terms to our model. Our Langevin equation in Eq.~(\ref{langevin}) gets just an additional term on the right-hand side, namely $\hbar \omega_D A_D a^*(t)/4$, and $\hbar \omega_D A_D a(t)/4$ for the complex conjugate equation for $a^*(t)$. These will then also have to enter into the streaming terms of the Fokker-Planck equation, leading to 
\begin{widetext}
\begin{dmath} \label{fpwhiteanddrive}
\begin{aligned}
    i \hbar \frac{\partial}{\partial t} P[a^*,a;t] = & -\frac{\partial}{\partial a} \bigg \{ (-\hbar \delta + \hbar g |a|^2) a P[a^*,a;t] + \frac{\hbar \omega_D A_D}{4} a^* P[a^*,a;t] \bigg \} \\
    & + \frac{\partial}{\partial a^*} \bigg \{ (-\hbar \delta + \hbar g^{*} |a|^2) a^* P[a^*,a;t] + \frac{\hbar \omega_D A_D}{4} a P[a^*,a;t] \bigg \} \\
    &- \frac{\hbar g^{K}}{2} \frac{\partial}{\partial a^*} \bigg \{ |a|^2 \frac{\partial}{\partial a} P[a^*,a;t] \bigg \}  - \frac{\hbar g^{K}}{2} \frac{\partial}{\partial a} \bigg \{ |a|^2 \frac{\partial}{\partial a^*} P[a^*,a;t] \bigg \}.
    \end{aligned}
\end{dmath}
\end{widetext}
This is now the complete Fokker-Planck equation, containing all the non-equilibrium dynamics of the space-time crystal in the semiclassical approximation. As this equation accurately describes the prethermalization of the space-time crystal observed experimentally, it is the most important result of our paper.  

\begin{figure*}
 {\centering
     \begin{subfigure}[b]{0.4\textwidth}
     \centering
         \includegraphics[width=\textwidth]{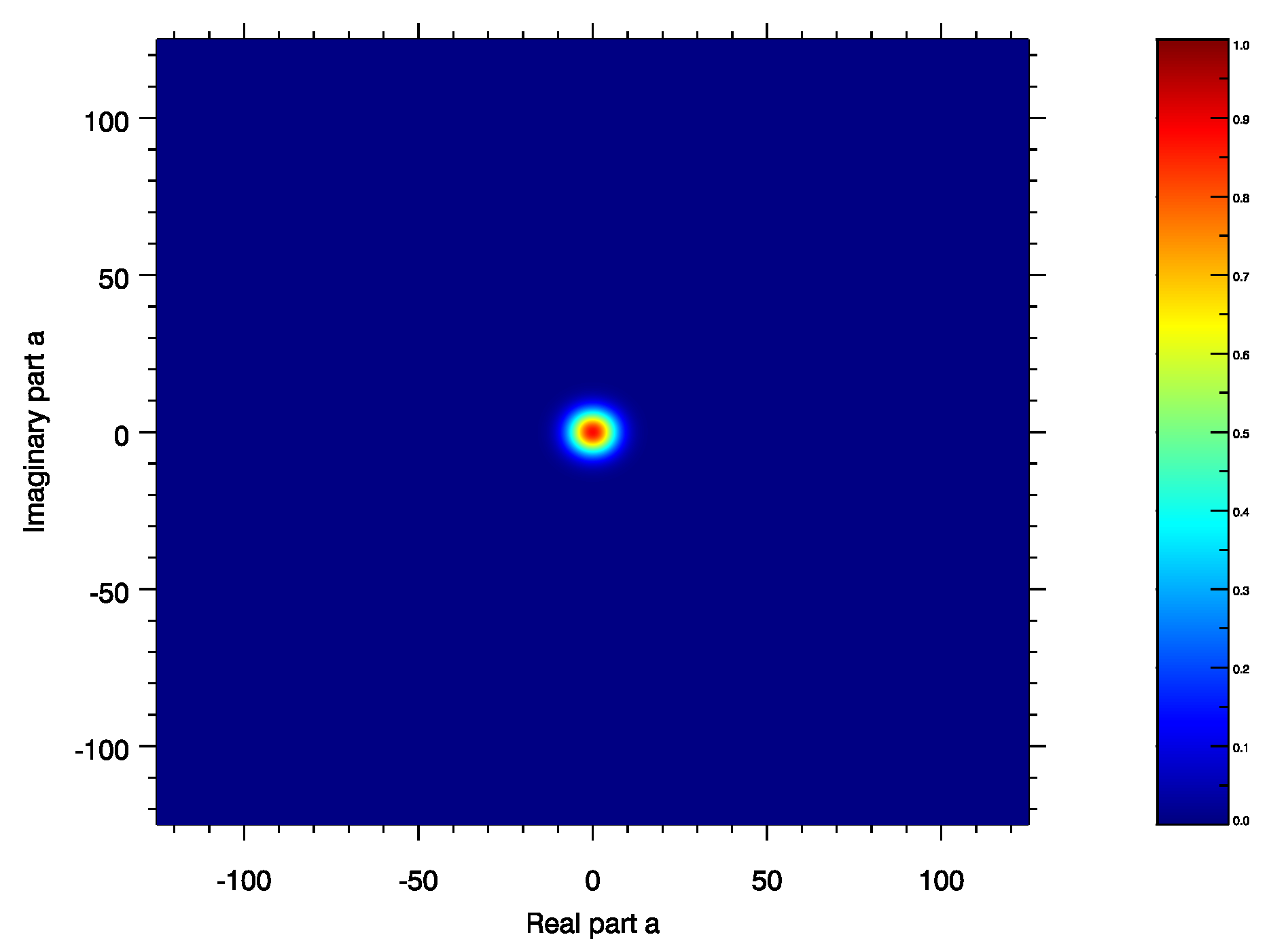}
         \caption{After 0 oscillations.}
     \end{subfigure}
     \begin{subfigure}[b]{0.4\textwidth}
         \centering
         \includegraphics[width=\textwidth]{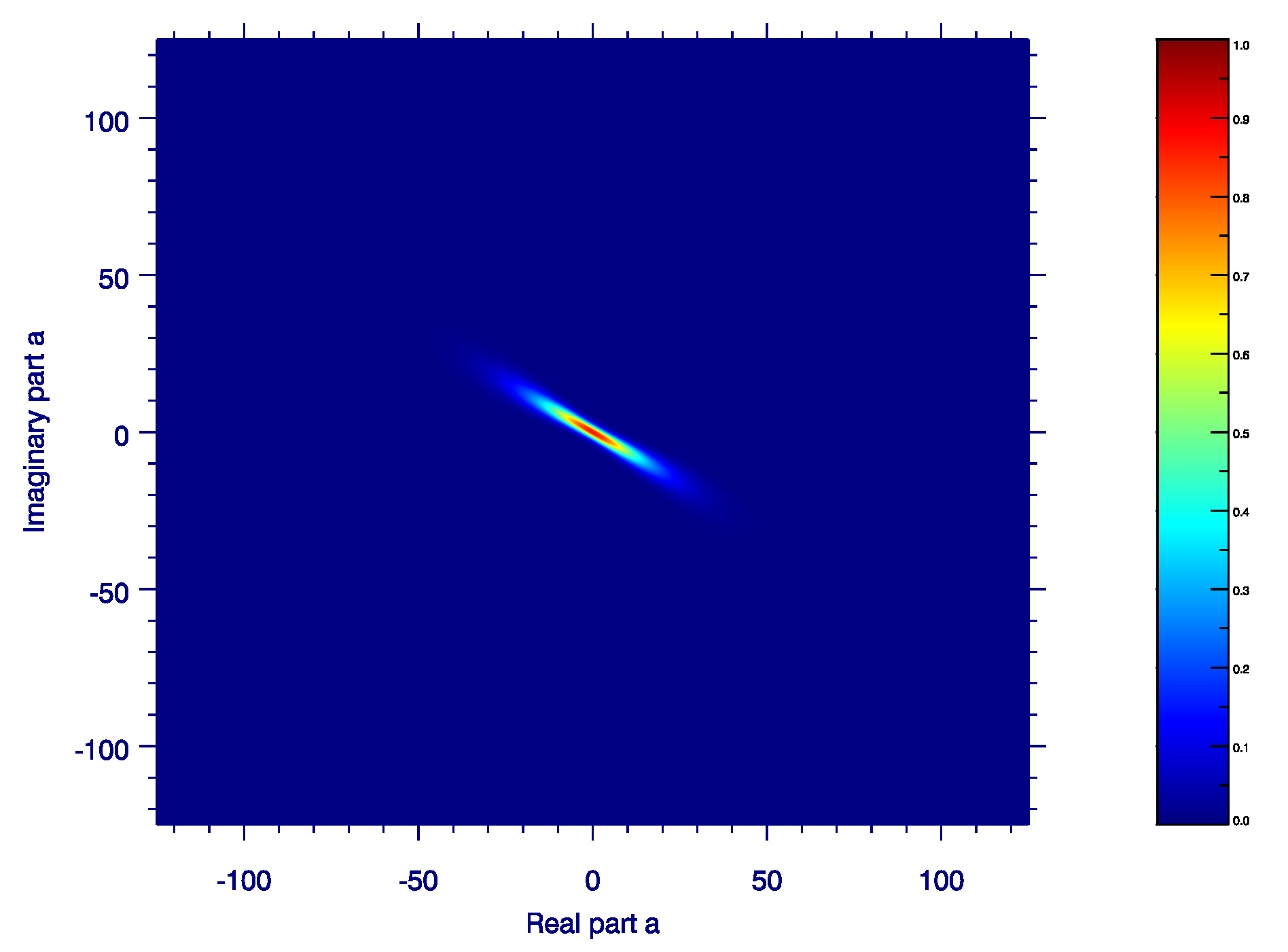}
         \caption{After 10 oscillations.}
     \end{subfigure}

     \begin{subfigure}[b]{0.4\textwidth}
         \centering
         \includegraphics[width=\textwidth]{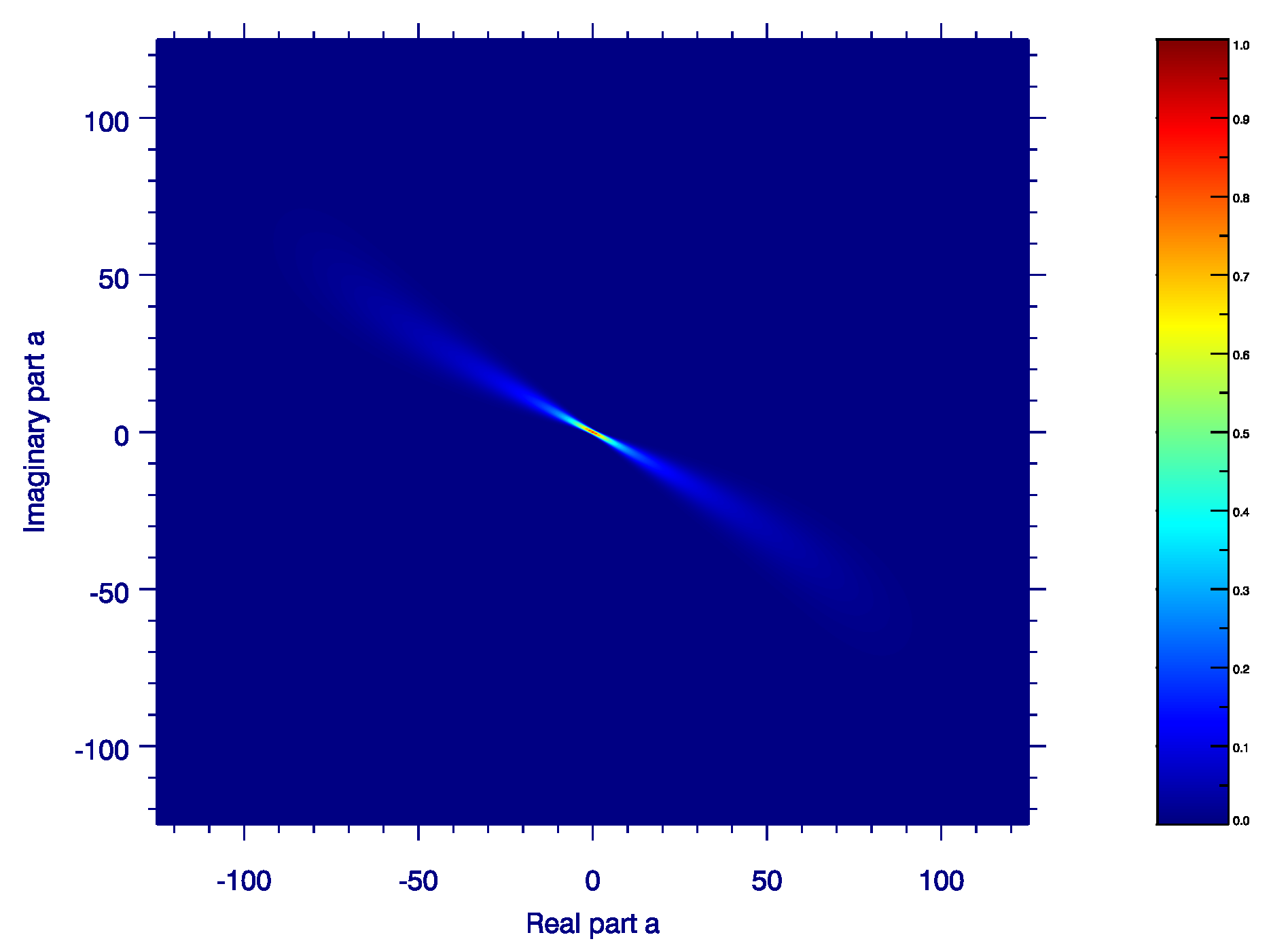}
         \caption{After 20 oscillations.}
     \end{subfigure}
     \begin{subfigure}[b]{0.4\textwidth}
     \centering
         \includegraphics[width=\textwidth]{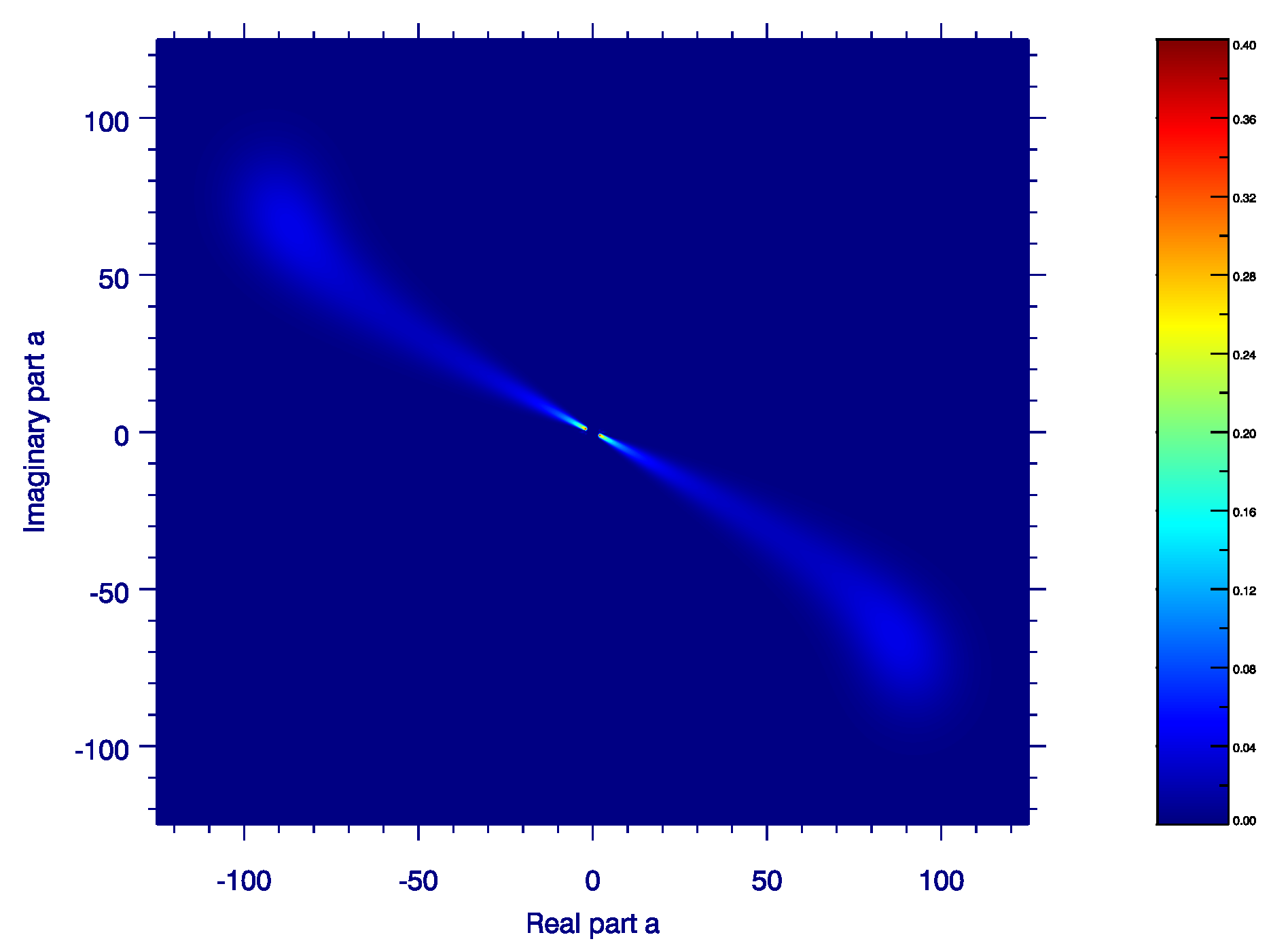}
         \caption{After 30 oscillations.}
     \end{subfigure}

     \begin{subfigure}[b]{0.4\textwidth}
         \centering
         \includegraphics[width=\textwidth]{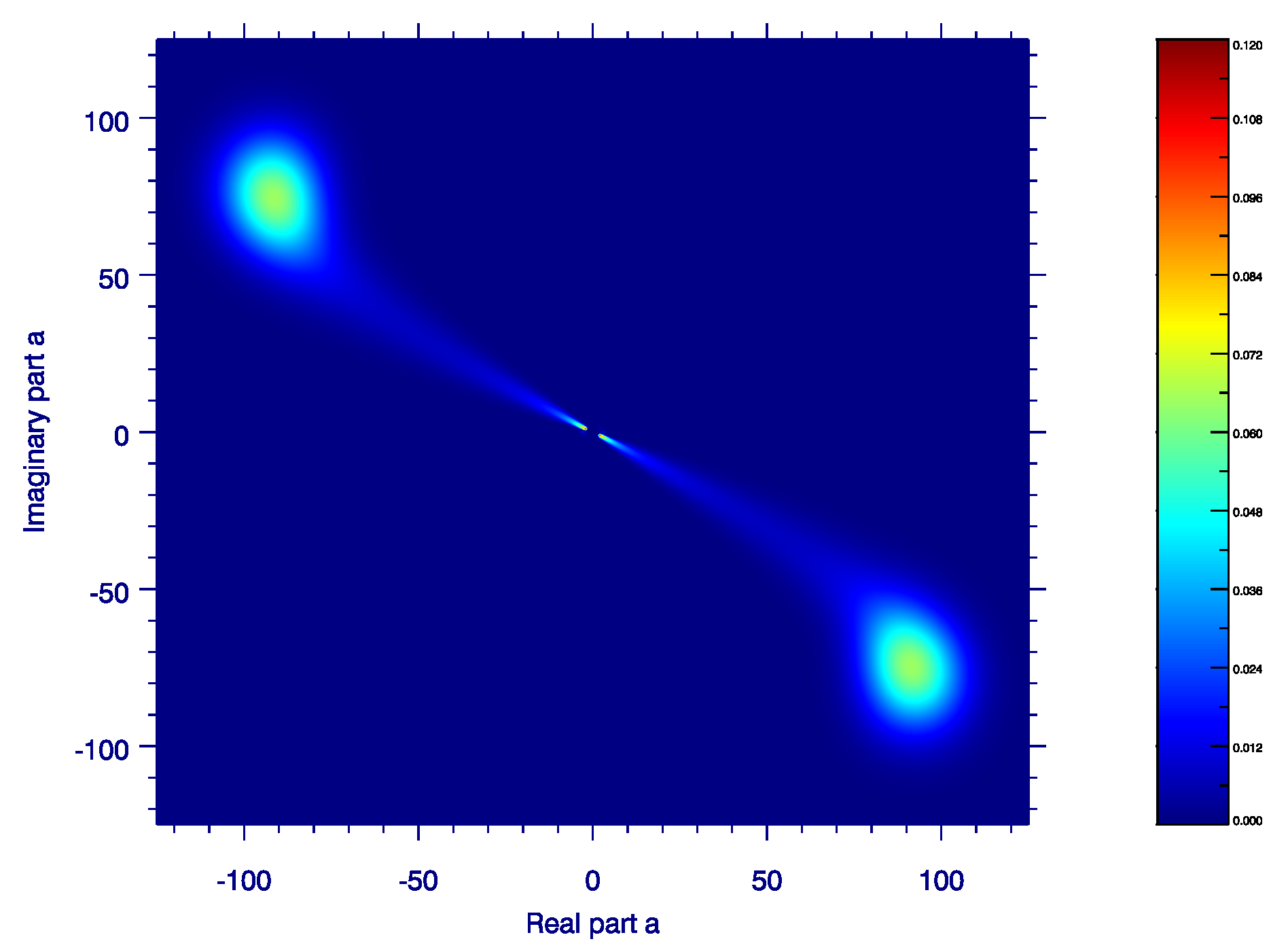}
         \caption{After 40 oscillations.}
     \end{subfigure}
     \begin{subfigure}[b]{0.4\textwidth}
         \centering
         \includegraphics[width=\textwidth]{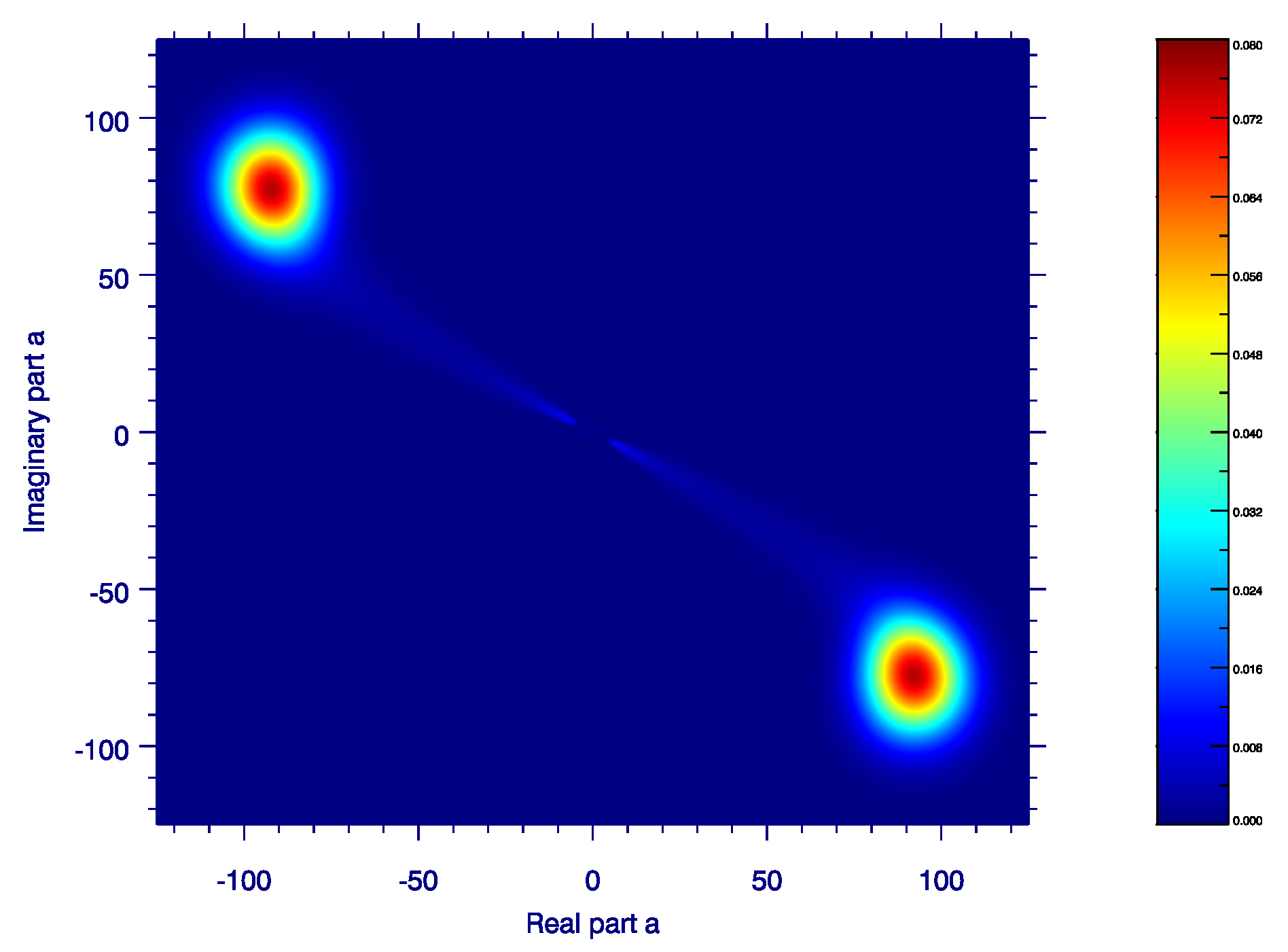}
         \caption{After 50 oscillations.}
     \end{subfigure}}
     \caption{Dynamical behavior of the probability distribution $P[a^*,a;t]$ in the complex $(a',a'')$ plane. Numerically, fifty oscillations, i.e., fifty periods of the drive, are simulated, and a picture was taken after every ten oscillations. The number of oscillations past is indicated below each figure. For clarity, we have suppressed the center part of the distribution after 20 oscillations, since otherwise the range of the color scale is suppressed by one order of magnitude.}
        \label{fig:trajectories}
\end{figure*} 

\subsubsection{Numerical solution} \label{numerics}

We have numerically solved the Fokker-Planck equation in Eq.~(\ref{fpwhiteanddrive}) by using an implicit finite-difference method, see App.~\ref{num-sim} for details. We have used the same parameters as obtained experimentally in Ref.~\cite{exp}, namely $\delta/2\pi$ = 2 Hz, $|g|/2\pi$ = 3$\times 10^{-4}$ Hz, $\phi_g$ = -0.4$\pi$, and $A_D$ = 0.091, except that we have increased $|g|$ by a factor of 10 in order to decrease $|\langle a \rangle|^2$ by the same factor for numerical convenience. In particular, we simulate what happens when we start with the thermal Gaussian initial distribution in Eq.~(\ref{gaussdistribution}) and then let the system evolve in time under influence of the drive. Note that our initial condition is $U(1)$ invariant, i.e., rotationally invariant in the complex $(a', a'')$ plane, and therefore does not contain a preferred phase. It is also invariant under inversion with respect to the origin $|a|=0$, which is the $Z_2$ symmetry we are most interested in. So the initial condition does not break any symmetries as desired for a discussion of spontaneous symmetry breaking. Our numerical findings of the spontaneous symmetry breaking in our time crystal is shown in Fig.~\ref{fig:trajectories}. We clearly see that initially the drive is only squeezing the probability distribution, which breaks the initial $U(1)$ symmetry down to a $Z_2$ symmetry, as only the inversion symmetry in the complex plane remains. This symmetry breaking is not spontaneous as the drive is explicitly breaking the $U(1)/Z_2$ symmetry. In this stage of the evolution the probability distribution is squeezed, but still has a maximum in the origin of the complex plane. In terms of order parameters this phase is thus characterized by $\langle a a \rangle \neq 0$ and $\langle a \rangle =0$, so the $Z_2$ symmetry is unbroken. Only at a later stage this single maximum in the origin falls apart into two maxima away from the origin that are related by the $Z_2$ inversion symmetry. 

\begin{figure}[b]
 {\centering 	\includegraphics[width=0.49\textwidth]{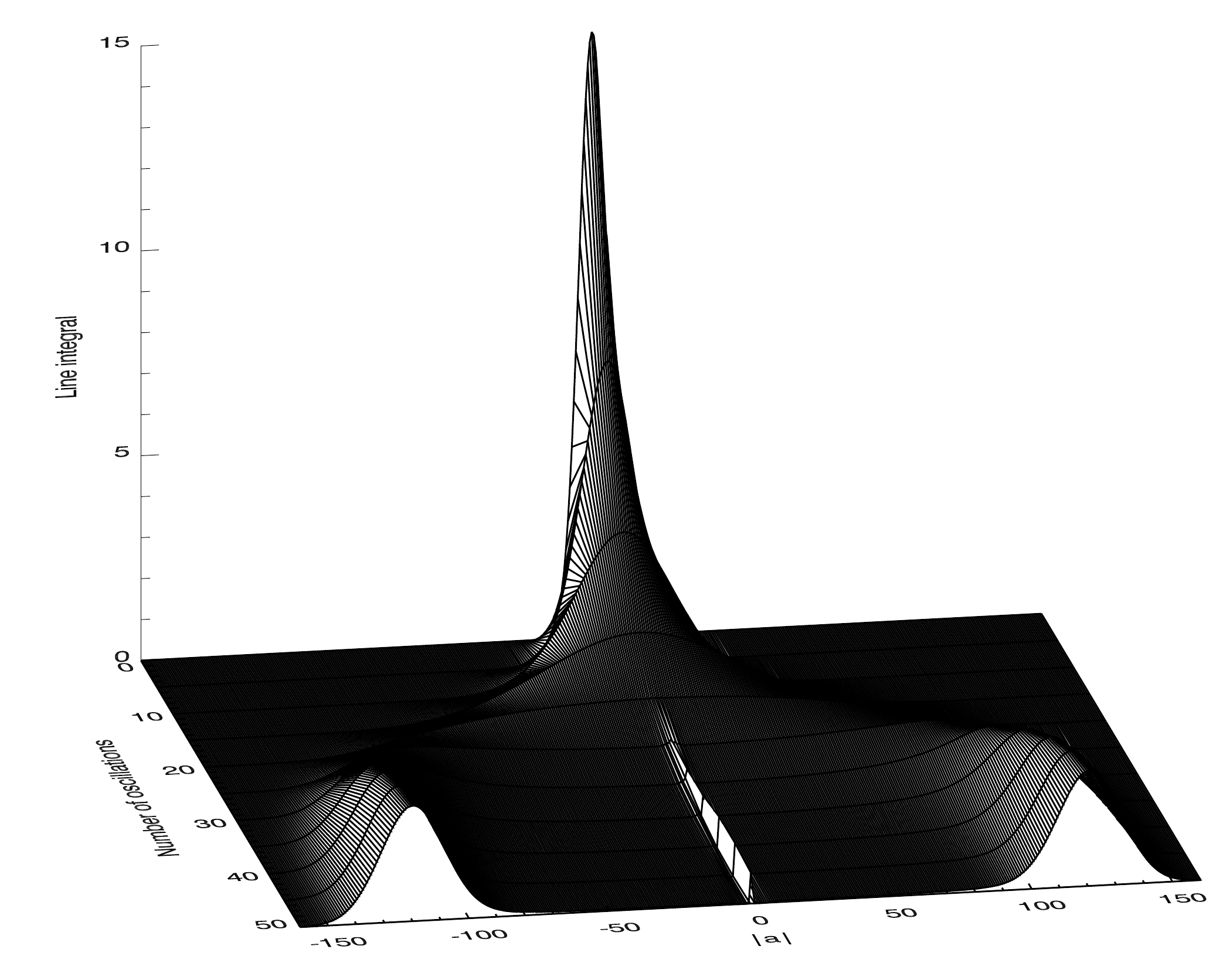}}
	\caption{Time evolution of the reduced probability distribution on the $\phi = 3\pi/4$ (mod $\pi$) diagonal, obtained by performing the integral of the probability distribution $P[a^*,a;t]$ over all lines perpendicular to this diagonal. }
        \label{fig:sommen}
\end{figure} 

To emphasize this effect, we have integrated the probability distribution over all lines perpendicular to the $\phi = 3\pi/4$ (mod $\pi$) diagonal and show the results for this reduced probability distribution in Fig.~\ref{fig:sommen}. Initially, the reduced distribution is Gaussian with its center at $|a|$ = 0. After the drive has been turned on, the central part of the distribution spreads out to the left and right and after 20 oscillations two distinct peaks appear. However, note that even for a large number of oscillations there remains a small part of the distribution near $a$ = 0. This comes about because the strength of the driving $\hbar \omega_D A_D/4$ is multiplied with $a$ and $a^*$ in the Fokker-Planck equation and so for $|a| \simeq 0$ the effect of the drive becomes negligible. Moreover the noise is multiplicative and vanishes in this limit. 

The appearance of the two symmetry-related peaks corresponds to the true spontaneous breaking of the $Z_2$ symmetry and the formation of the space-time crystal as now both $\langle a a \rangle \neq 0$ and $\langle a \rangle \neq 0$. Note that in principle the probability distribution is inversion symmetric at all times, as the Hamiltonian does not break this symmetric explicitly. This implies that if one performs many experiments with the same thermal initial condition for the Bose-Einstein condensate, the observed space-time crystal will have a temporal phase of $\phi$ or $\phi + \pi$ with exactly equal probability \cite{ssb}.

\subsection{Quantum dynamics} \label{colourednoise}

In classical non-equilibrium physics we usually deal with white noise, and a frequency-independent fluctuation-dissipation theorem. Perhaps the most famous example of white noise is Brownian motion with the Einstein relation between the diffusion constant and temperature. In the above semi-classical approach for the experimentally relevant (prethermalization) time scales, the same approximation was made by evaluating the exact fluctuation-dissipation theorem in Eq.~(\ref{fltheorem}) at twice the typical frequency $\bar{\omega}$. Quantum noise, however, is in general colored noise and this complicates the analytical or numerical analysis of the relevant Langevin problem considerable. Fortunately, if we are interested in the longest time scales of the problem in which ultimately relaxation towards thermal equilibrium occurs, we are allowed to make another approximation, i.e., we can take the long-wavelength or in this case the low-frequency limit. In that limit $g^K(\omega)$ goes to the nonzero constant $g^K(0)$ and we have 
\begin{equation}
    g^K(t-t') \simeq g^K(0) \delta(t-t'),
\end{equation}
which again corresponds to white noise in frequency space. We would like to take the same limit in all quantities that appear in the fluctuation-dissipation theorem, but this is not possible because we have
\begin{dmath}
    (1+2N(\hbar \omega)) \simeq \frac{2 k_B T}{\hbar \omega},
\end{dmath}
which in fact diverges for $\omega \rightarrow 0$. In order to compensate for this divergence in the exact fluctuation-dissipation theorem in Eq.~(\ref{fltheorem}), we therefore obtain that in frequency space we must have that $g''(\omega) = (\hbar \omega/4ik_B T) g^K(0)$ or in the time domain that
\begin{dmath}
    g''(t - t') = \frac{g^K(0)}{4k_B T} \hbar \frac{\partial}{\partial t} \delta (t-t').
\end{dmath}

\subsubsection{Langevin formulation} \label{L formulation2}

With that, our Langevin equation in Eq.~(\ref{langevin}) becomes
\begin{eqnarray} \label{langevingeneraltocoloured}
    \bigg ( 1 &-& \frac{\hbar g^K}{2 k_B T} a^*(t) a(t) \bigg ) i\hbar \frac{\partial}{\partial t} a(t) \\ 
    &=& -\hbar \delta a(t) + \hbar g' a^*(t) a(t) a(t)
   + \eta(t) a^*(t), \nonumber
\end{eqnarray}
and the complex conjugate equation for $a^*(t)$. Notice that for $g'$ we are allowed to take the naive zero-frequency limit and that from now on we use $g^K \equiv g^K(\omega=0)$. Moreover, for the same reasons as before we have briefly returned to the case without drive, as the unitary evolution of the drive is easily incorporated later on as we have seen. We also introduce the (stochastic) quantities
\begin{dmath}
    I(t)  = 1 - \frac{\hbar g^K}{2 k_B T} |a(t)|^2
\end{dmath} 
and thus
\begin{dmath}
    I^*(t) = 1 + \frac{\hbar g^K}{2k_B T} |a(t)|^2,
\end{dmath}
as $g^K$ is purely imaginary. These will play a key role in our following discussion. 

We now want to work out again the relevant correlation functions for the noise $\eta(t)$ with certain functions of $a(t)$ and $a^*(t)$, in order to find the equations of motion for $\langle a(t) \rangle$ and $\langle a^*(t) a(t) \rangle$, similar to Eqs.~(\ref{langevinwhite}) and (\ref{eomwhite}). By dividing the Langevin equation in Eq.~(\ref{langevingeneraltocoloured}) by $I(t)$, we can see that our task now is to calculate quantities like $\langle \eta(t) a^*(t)/I(t) \rangle$ and $\langle \eta(t) a^*(t) a^*(t)/I(t) \rangle$. Let us therefore also define 
\begin{dmath}
    A^*(t) \equiv \frac{a^*(t)}{I(t)}.
\end{dmath}
We can expand this function as
\begin{dmath}
    A^*(t) = A^*(0) + \frac{\partial A^*(0)}{\partial a^*} (a^*(t)-a^*(0)) + \frac{\partial A^*(0)}{\partial a} (a(t)-a(0)), 
\end{dmath}
neglecting all higher-order contribution. So we always need to remember that these stochastic quantities we introduced are explicit functions of $a^*(t)$ and $a(t)$.

Using again that $\langle \eta(t) \rangle = \langle \eta^*(t) \rangle = \langle \eta(t) \eta(t') \rangle = \langle \eta^*(t) \eta^*(t') \rangle = 0$, and the general solution for $a(t)$ in this case given by
\begin{dmath}
      a(t) = a(0) - \frac{i}{\hbar} \int_0^t dt' \bigg \{ - \frac{\hbar\delta}{I(t')} a(t') + \hbar g' A^*(t') a(t')a(t') +  \eta(t')A^*(t') \bigg \},
\end{dmath}
we are in the position to calculate the desired correlation functions. The results become
\begin{equation}
\langle \eta(t) A^*(t) \rangle =  - \frac{\hbar g^K}{2} \left\langle \frac{\partial A^*(t)}{\partial a^*} A(t) \right\rangle ,
\end{equation}
and
\begin{eqnarray}
\langle \eta(t) a^*(t) A^*(t) \rangle &=&  - \frac{\hbar g^K}{2} \left\langle A(t) A^*(t) \right\rangle  \nonumber \\
&-& \frac{\hbar g^K}{2} \left\langle a^*(t) \frac{\partial A^*(t)}{\partial a^*} A(t) \right\rangle.
\end{eqnarray}
With that, the full set of equations of motion are now obtained as
\begin{dmath}\label{langevincoloured}
   i\hbar \frac{\partial}{\partial t} \langle a(t) \rangle = \bigg \langle (-\hbar \delta + \hbar g' a^*(t) a(t)) \frac{a(t)}{I(t)} \bigg \rangle - \frac{\hbar g^K}{2} \bigg \langle \frac{\partial A^*(t)}{\partial a^*} A(t) \bigg  \rangle,
\end{dmath}
the complex conjugate equation for $\langle a^*(t) \rangle$, and 
\begin{eqnarray}\label{eomcoloured}
  && \hspace{-0.5cm} i\hbar\frac{\partial}{\partial t} \langle a^*(t) a(t) \rangle  \\
  && = \bigg \langle (-\hbar \delta + \hbar g' a^*(t) a(t)) \bigg (\frac{a(t)}{I(t)} - \frac{a^*(t)}{I^*(t)} \bigg ) \bigg \rangle \nonumber \\
  && - \frac{\hbar g^K}{2} \left\langle a^*(t) \frac{\partial A^*(t)}{\partial a^*} A(t) + a(t) \frac{\partial A(t)}{\partial a} A^*(t) \right\rangle \nonumber \\
  && - \hbar g^K \langle A^*(t) A(t) \rangle , \nonumber
\end{eqnarray}
which may be directly compared to Eqs.~(\ref{langevinwhite}) and (\ref{eomwhite}).
\vspace{0.3cm}

\subsubsection{Fokker-Planck formulation} \label{FP formulation2}

Following the same line of thought as in the semi-classical case, we obtain for the case without drive the Fokker-Planck equation
\begin{eqnarray} \label{fpcoloured}
i\hbar \frac{\partial}{\partial t} P[a^*,a;t] &=& -\frac{\partial}{\partial a} \bigg\{ (-\hbar \delta + \hbar g' |a|^2) \frac{a}{I} P[a^*,a;t] \bigg \} \nonumber \\
&+& \frac{\partial}{\partial a^*} \bigg\{ (-\hbar \delta + \hbar g' |a|^2) \frac{a^*}{I^*}  P[a^*,a;t] \bigg \} \nonumber \\
&-& \frac{\hbar g^{K}}{2} \frac{\partial}{\partial a^*} \bigg \{ \frac{a}{I^*} \frac{\partial}{\partial a}  \frac{a^*}{I} P[a^*,a;t] \bigg \} \nonumber \\
&-& \frac{\hbar g^K}{2} \frac{\partial}{\partial a} \bigg \{ \frac{a^*}{I} \frac{\partial}{\partial a^*} \frac{a}{I^*} P[a^*,a,t] \bigg \} ,
\end{eqnarray}
where we want to emphasize the fact that $g$ and $g^*$ in Eq.~(\ref{fpwhite}) have now both become equal to $g'$. As a result the dissipation in the equations of motion effectively does no longer enter via the nonlinear coupling constant, but through the imaginary parts of the quantities $I$ and $I^*$ instead. 

This Fokker-Planck equation can be derived directly from the effective action in Eq.~(\ref{action}) in the same manner as described before. Instead of a constant $g''$ in Eq.~(\ref{fpdirectaction}), we now have a contribution containing a time derivative. For this situation, the canonical momentum $\pi(t)$ therefore reads
\begin{dmath}
    \pi(t) = \frac{i\hbar I(t)}{\hbar g^K |a(t)|^2}
    \bigg ( -i\hbar \frac{\partial}{\partial t} +\hbar \delta -\hbar g' |a(t)|^2 \bigg )  a^*(t), 
\end{dmath}
and the Hamiltonian becomes after performing the usual Legendre transformation
\begin{eqnarray}
    H(t) &=&  \frac{\pi(t)}{i\hbar I(t)} (-\hbar \delta + \hbar g'|a(t)|^2)a(t) \\ &-& \frac{\pi^*(t)}{i\hbar I^*(t)}(-\hbar \delta +\hbar g'|a(t)|^2)a^*(t) -  \frac{g^K |a(t)|^2}{\hbar |I(t)|^2} \lvert \pi(t)\rvert ^2, \nonumber
\end{eqnarray}
which after quantization, and resolution of the usual operator-ordering problems, indeed exactly reproduces the Fokker-Planck equation in Eq.~(\ref{fpcoloured}). 

If we also include the drive terms again, similar to the approach in the semi-classical case, the full Fokker-Planck equation reads now
\begin{widetext}
\begin{dmath} \label{fpcolouredanddrive}
\begin{aligned}
i\hbar \frac{\partial}{\partial t} P[a^*,a;t] = & -\frac{\partial}{\partial a} \bigg \{ (-\hbar \delta + \hbar g' |a|^2) \frac{a}{I} P[a^*,a;t] + \frac{\hbar \omega_D A_D}{4} \frac{a^*}{I} P[a^*,a;t] \bigg \} \\
& + \frac{\partial}{\partial a^*}\bigg\{ (-\hbar \delta + \hbar g' |a|^2) \frac{a^*}{I^*} P[a^*,a;t] + \frac{\hbar \omega_D A_D}{4} \frac{a}{I^*} P[a^*,a;t] \bigg \} \\
& - \frac{\hbar g^{K}}{2} \frac{\partial}{\partial a^*} \bigg \{ \frac{a}{I^*} \frac{\partial}{\partial a} \frac{a^*}{I} P[a^*,a;t] \bigg \} 
- \frac{\hbar g^K}{2}  \frac{\partial}{\partial a} \bigg \{ \frac{a^*}{I} \frac{\partial}{\partial a^*} \frac{a}{I^*} P[a^*,a;t] \bigg \} .
\end{aligned}
\end{dmath}
Although this appears to be more complicated then in the semi-classical case, the equilibrium distribution can now be obtained analytically and ultimately reads
\begin{dmath} \label{quantumequil}
    P[a^*,a] \propto |I(|a|^2)|
    \exp \bigg ( - \frac{1}{k_B T}  \bigg \{ -\hbar \delta |a|^2 +\frac{\hbar g'}{2}|a|^4 + \frac{\hbar \omega_D A_D}{8} (a^* a^* + a a) \bigg \} \bigg ).
\end{dmath}
\end{widetext}
Notice that the equilibrium distribution scales with the Boltzmann factor $e^{- E/k_B T}$, where $E[a^*,a]$ is the classical energy associated with the quantum Hamiltonian in Eq.~(\ref{Hamiltonian}), which was the starting point of our developments. Apart from this Boltzmann factor, we find also a prefactor equal to $|I(|a|^2)|$. This precisely corresponds to the expected change in phase-space volume \cite{niu, achim} as our Hamiltonian dynamics is modified into
\begin{equation}
i\hbar I \frac{\partial a}{\partial t} = \frac{\partial E[a^*,a]}{\partial a^*},~
-i\hbar I^* \frac{\partial a^*}{\partial t} =
\frac{\partial E[a^*,a]}{\partial a} .
\end{equation}
In the overdamped limit $\hbar |g^K| \gg k_B T$ this prefactor becomes $\hbar |g^K| |a|^2/2k_B T$ and therefore the stationary solution will always be zero at $|a| = 0$. On the other hand, in the underdamped limit, we have $I \simeq 1$ and the prefactor doesn't depend on $a$ or $a^*$ and thus it can be absorbed in the normalisation constant. For the experimental parameters of Ref.~\cite{exp}, the underdamped limit is indeed a very good approximation, which confirms that dissipation is only playing a minor role in this case as expected for a superfluid droplet that is almost completely Bose-Einstein condensed.

\section{Comparison with experiments} \label{experiments}

\begin{figure*}
 {\centering
     \begin{subfigure}[b]{0.4\textwidth}
     \centering
         \includegraphics[width=\textwidth]{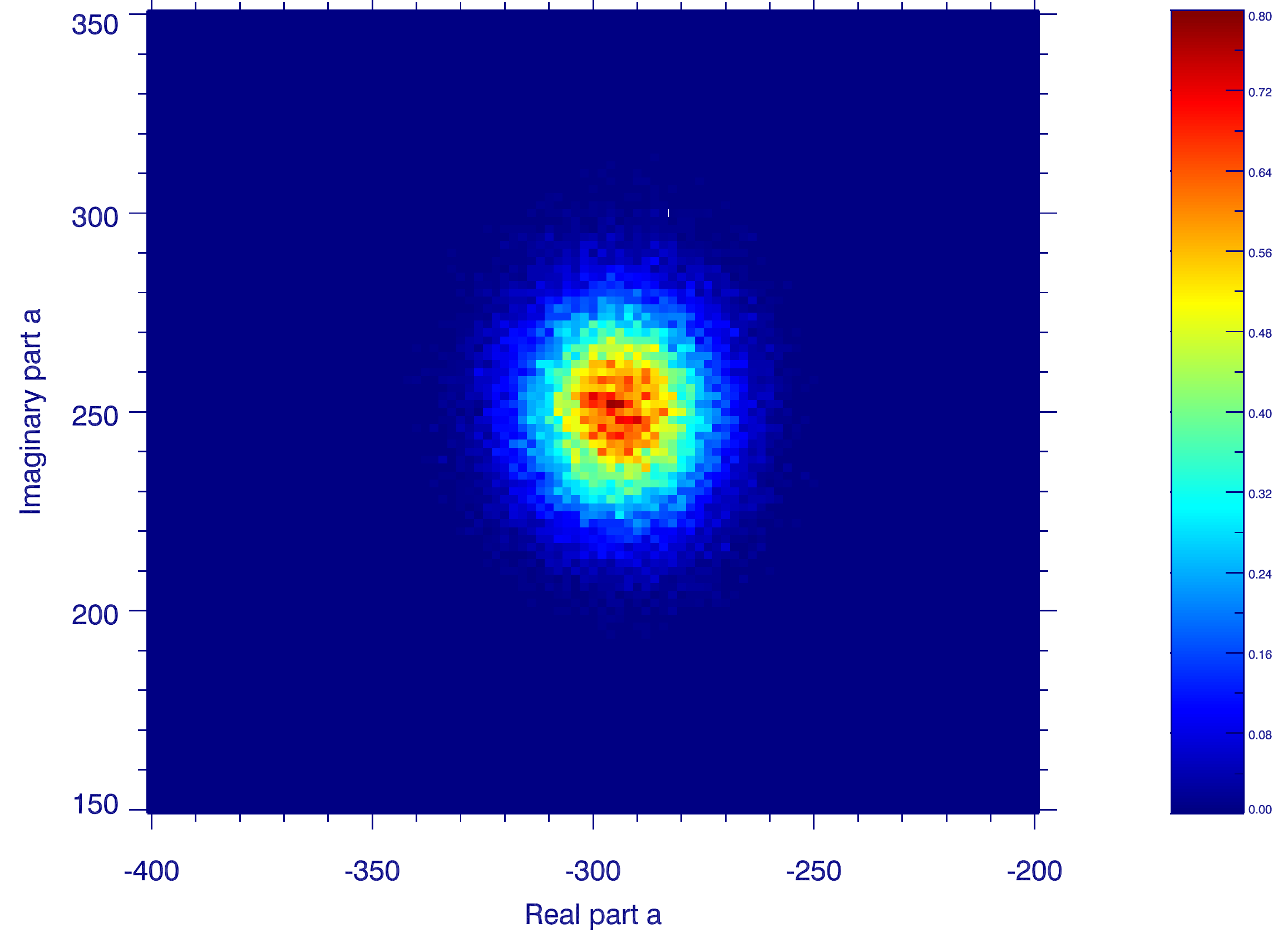}
         \caption{Equilibrium distribution for the semi-classical approach using the Langevin equation.}
     \end{subfigure}
     \begin{subfigure}[b]{0.4\textwidth}
         \centering
         \includegraphics[width=\textwidth]{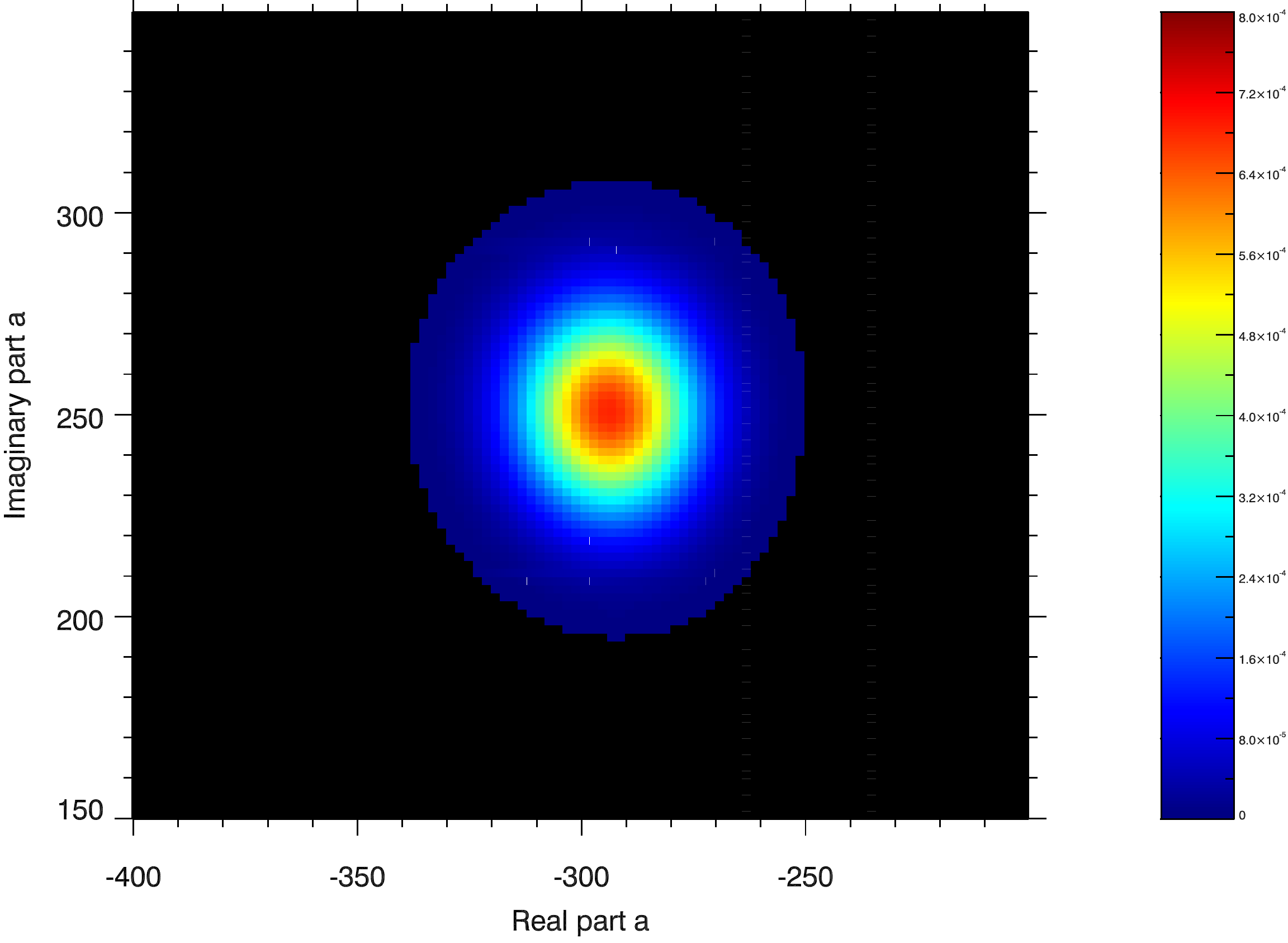}
         \caption{Equilibrium distribution for the semi-classical approach using the Fokker-Planck equation.}
     \end{subfigure}

     \begin{subfigure}[b]{0.4\textwidth}
         \centering
         \includegraphics[width=\textwidth]{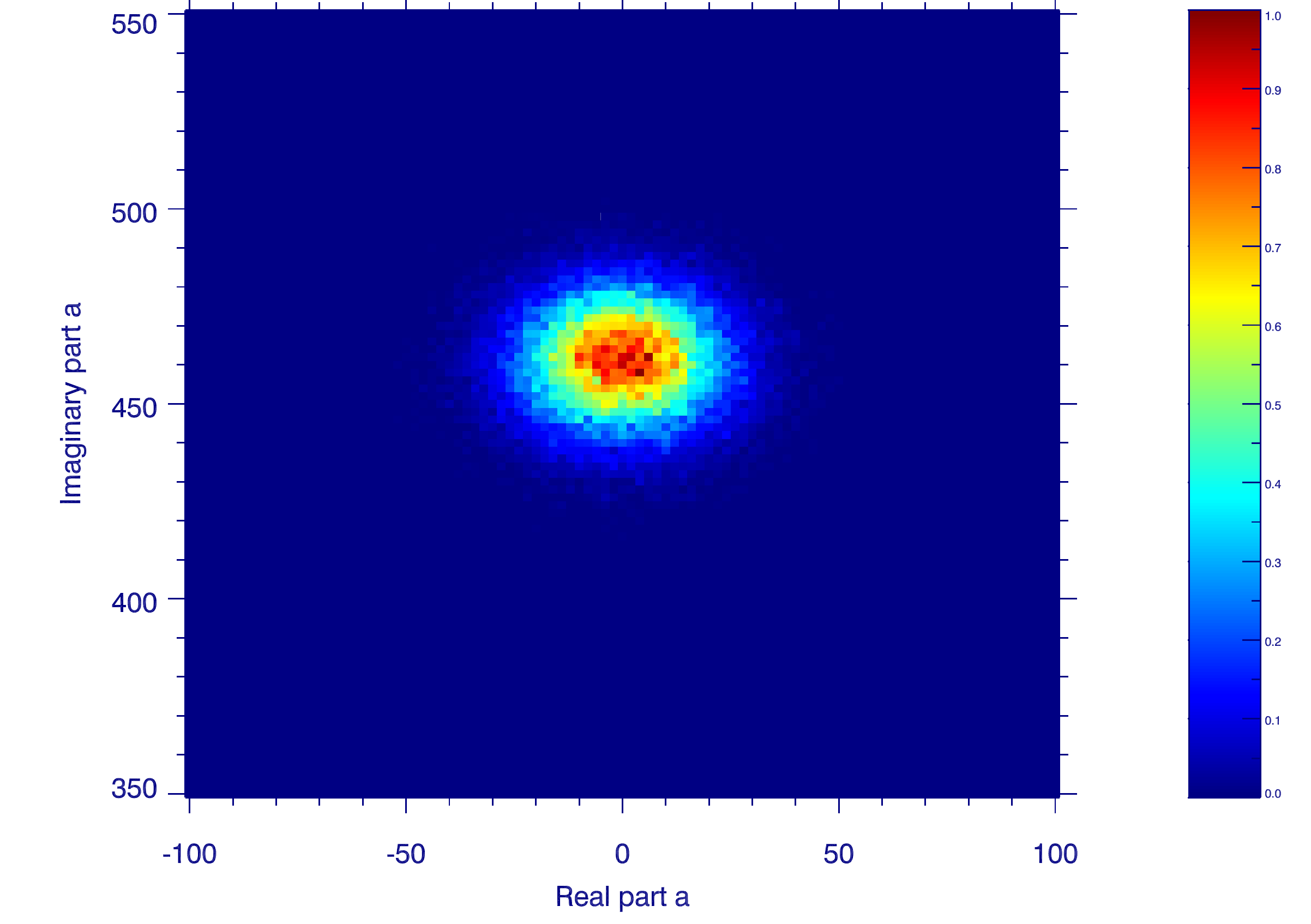}
         \caption{Equilibrium distribution for the quantum approach using the Langevin equation.}
     \end{subfigure}
     \begin{subfigure}[b]{0.4\textwidth}
     \centering
         \includegraphics[width=\textwidth]{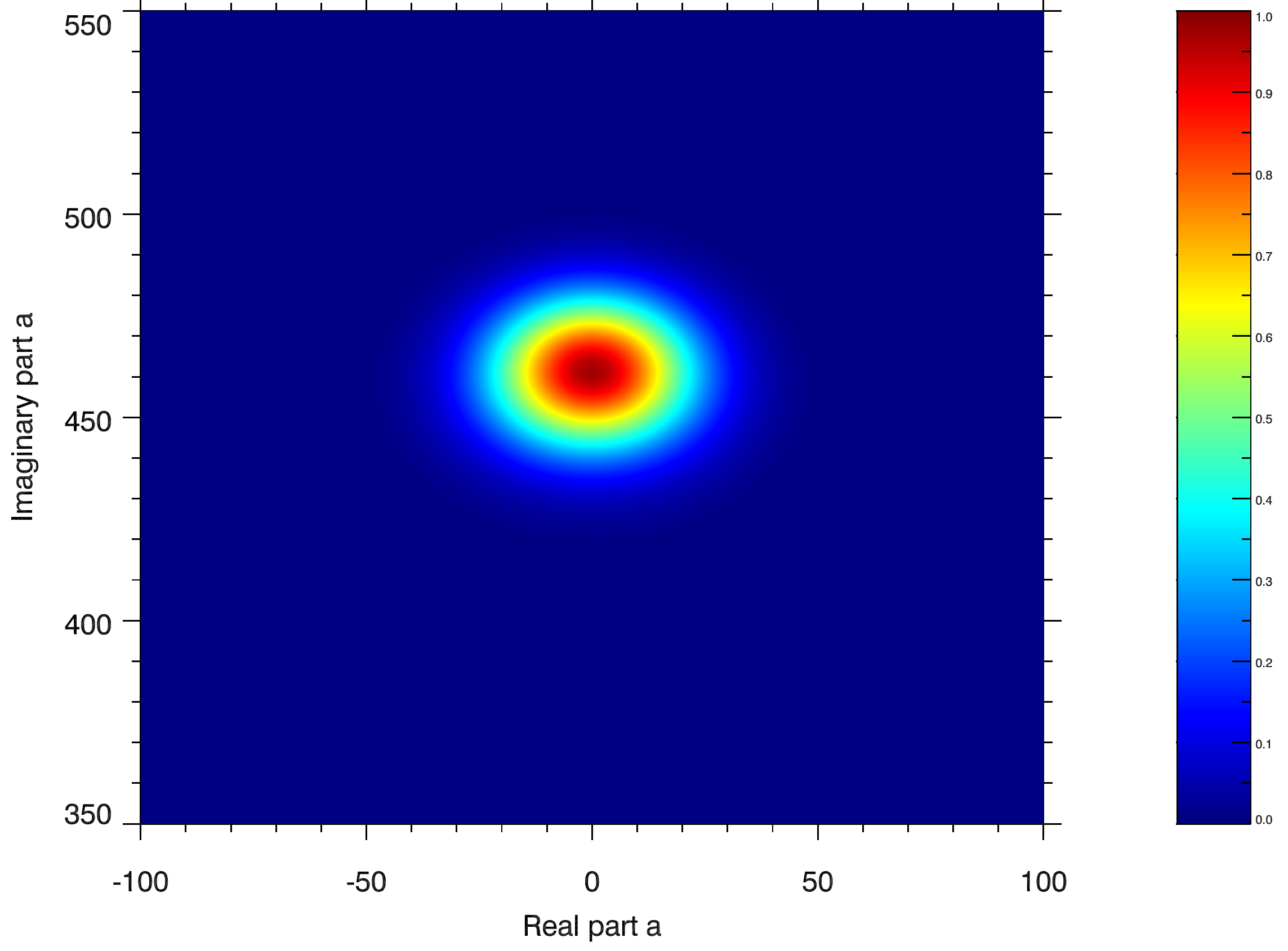}
         \caption{Equilibrium distribution for the quantum approach using the Fokker-Planck equation.}
     \end{subfigure}}

     \caption{Comparison between the two numerical simulations of the equilibrium probability distributions $P[a^*,a]$ for both the semi-classical and quantum cases, respectively.  All the results are calculated for the experimental parameters of Ref.~\cite{exp}.}
        \label{fig:equilibrium}
\end{figure*}

In this section, our goal is to show the result for the stationary solution of the Fokker-Planck equation for the quantum dynamics in Eq.~(\ref{quantumequil}) and compare it with the semi-classical dynamics, either using the Langevin equation or the Fokker-Planck equation. The semi-classical dynamics within the Langevin formulation, after including also technical noise, is found to agree very well with experiments in Ref.~\cite{ssb}. In Fig.~\ref{fig:equilibrium} we therefore compare the result for the equilibrium distribution from the Langevin equation with the result from the Fokker-Planck equation, both for the semi-classical approach, and obtain, as expected, excellent agreement. The results for the Langevin equation are obtained by calculating 100,000 trajectories using random values of $a$ and $a^*$ from the initial distribution of Eq.~(\ref{gaussdistribution}) and integrating the evolution of the individual trajectories over 100 oscillations, as discussed in Ref.~\cite{ssb}. The results are then binned to obtain the probability distribution $P[a^*,a]$. The results for the Fokker-Planck distribution are obtained by taking an initial Gaussian distribution and letting it evolve to an equilibrium using Eq.~(\ref{fpwhiteanddrive}).  Not only do the two equations lead to the same equilibrium, also the dynamics towards the equilibrium is very similar in both cases and this numerically validates the equivalence between the two theoretical frameworks.

To compare the semi-classical dynamics to the quantum dynamics, the Langevin equation of Eq.~(\ref{langevingeneraltocoloured}) is used to obtain the stochastic trajectories in the quantum approach. Since in this approach the imaginary part $g''$ of the interaction parameter has effectively been removed from the equations of motion in favor of $g^K$, we use $g'=|g|$ with $|g|$ the interaction parameter as previously used in the semi-classical approach. In detail the probability distributions are very different as implicitly also found in Ref.~\cite{exp}. One important difference is immediately clear from the location of the peaks. In the semi-classical simulations, the peaks are located almost on the diagonal with $\phi \simeq 3\pi/4~({\rm mod}~\pi)$, whereas in the ultimate equilibrium state the peaks are located on the imaginary axis. In the experiment we find the peak to locate nearly on the real axis, but in the experiment we have determined the amplitude of the highly-excited axial mode, and this amplitude is proportional to the time derivative of $a$ in the laboratory frame and this leads to a phase shift of $\pi/4$. Taking this subtle point into account, we obtain excellent agreement between the experiment and the simulations of the semi-classical Langevin equation as shown in Ref.~\cite{exp}. The location of the maxima in the semi-classical equilibrium also depends sensitively on the detuning used, whereas in the true equilibrium the peaks are always on the imaginary axis independent of detuning. For instance, in our numerical solution of the Fokker-Planck equation in Sec.~\ref{numerics} we used $g' \ll g''$. In that case we obtain from the equations of motion from $\langle a(t) \rangle$ that in the prethermal state
\begin{equation}
\cos(2\phi) = \frac{4\delta}{\omega_D A_D}~, \label{evenwicht_phi}
\end{equation}
with $\sin(2\phi)<0$, and
\begin{equation}
|\langle a \rangle|^2 = \frac{k_B T}{\hbar \bar{\omega}} 
  + \frac{1}{|g|} \sqrt{ \bigg( \frac{\omega_D A_D}{4} \bigg)^2 - \delta^2 }~. \label{evenwicht_rho}
\end{equation}
The latter agrees with Ref.~\cite{exp} except for the first term in the right-hand side that represents the contribution from the noise-induced drift and that is small compared to the second term under the conditions of interest. In that regime we can actually take the limit $\omega_D A_D \gg {\rm max}(4|\delta|, |g|k_B T/\hbar\bar{\omega})$ to show, next to $|\langle a \rangle| \simeq \sqrt{\omega_D A_D/4|g|}$, that the phase of the space-time crystal becomes in general
\begin{equation}
\phi \simeq \frac{\pi - \phi_g}{2}~({\rm mod}~\pi),
\end{equation}
where $-\pi<\phi_g<0$ is the phase of $g$ \cite{ssb}.

We thus confirm here that our semi-classical approach can very accurately describe the experiments and that the experiments have explored up to now only the relatively short time scales in which prethermalization occurs, but not yet the longer time scales on which the final thermalization would take place. The latter may be very hard to realize in practice for two reasons. First, the dissipation in the experiment also affects the radial breathing mode that acts as a drive for the axial collective modes. As a result the amplitude of the drive is not constant and slowly decays at those long time scales. Second, our analysis is based on the Hamiltonian in Eq.~(1), which is valid in the rotating frame of the drive after neglecting the non-resonant terms. On the long time scales of interest for the thermalization, these non-resonant, and therefore oscillating terms will lead to heating that requires a further analysis outside the scope of our paper.

For completeness, we reiterate that when we look at the width of the peaks, we notice that our theoretical results are always narrower than the experimental data. This can be explained by the fact that in the experiments additional broadening occurs from the fluctuations in the initial conditions, most importantly the number of atoms in the Bose-Einstein condensate. As is shown in Ref.~\cite{ssb} this can be most easily accounted for in our theoretical modeling by allowing for some fluctuations in the detuning $\delta$.



\section{Conclusion and outlook}\label{conclusion}
The main achievements of our work can be summarized as follows. We developed the general theory to describe the non-equilibrium physics of the space-time crystal in an atomic Bose-Einstein condensate. Starting from a non-linear (cubic) coupling of the space-time crystal to a heat bath, we applied the tools of the Schwinger-Keldysh formalism to derive the general form of the Langevin equation. At that point, we distinguished two cases: a semi-classical approach, where the frequency dependence of the dissipation was neglected and a fully quantum approach, where this frequency dependency was included in the long wave-length limit. For both cases, we presented the full system of equations describing the dynamics: both in the Langevin formulation as well as in the Fokker-Planck formulation. We also showed the consistency of the two formulations. Furthermore, we were able to derive stationary solutions for both of these Fokker-Planck equations, which, as expected, turned out to scale with the Boltzmann factor $e^{-E/k_B T}$, where $E[a^*,a]$ is the appropriate energy for that specific situation.

By solving the semi-classical Fokker-Planck equation numerically, we were able to visualize very explicitly the dynamical features of the spontaneous breaking of the $Z_2$ symmetry and the associated formation of the space-time crystal. Since the semi-classical dynamics describes the prethermalization at relative short time scales, it can be directly compared with the experimental data. Excellent agreement between theory and experiment is obtained in this manner. The quantum method on the other hand makes it possible to discuss the ultimate thermalization of our theoretical model, but such long time scales are presently not observable experimentally and may be a topic of further investigations.

An open end in our discussion is the origin of the cubic coupling term between the space-time crystal and the heat bath. Although its theoretical implications work out nicely as these are mostly based on universal features such as the fluctuation-dissipation theorem, its precise (experimental) origin remains unclear. In particular, we would like to understand better which modes of the thermal cloud effectively represent the heat bath in our approach. In this manner it might be possible to derive also the strength of the dissipation and the noise from first principles. Nonetheless, our results shows that we have already obtained a solid theoretical description of the experiments, with minimal approximations, that underpins the spontaneous symmetry breaking occurring in these kind of time crystals.

An interesting area for further research will be to use our Fokker-Planck equation in order to study the phenomenon of `tunneling' from one energy minimum to the other. Although we did not look into this yet in our discussion, the non-equilibrium physics of this problem is also fully captured by our model and may also be in reach in future experiments with space-time crystals in Bose-Einstein condensates.

\appendix
\section{Semi-classical equilibrium without drive} \label{appA}
The equilibrium in the semi-classical approach has to satisfy the stationary Fokker-Planck equation for that case, which reads
\begin{eqnarray}
    0 &=& -\frac{\partial}{\partial a} \bigg\{ (-\hbar \delta + \hbar g |a|^2) a P[a^*,a] \bigg \} \nonumber \\
    &+& \frac{\partial}{\partial a^*} \bigg \{ (-\hbar \delta + \hbar g^{*} |a|^2) a^* P[a^*,a] \bigg \} \nonumber \\
    &-& \frac{\hbar g^{K}}{2} \frac{\partial}{\partial a^*} \bigg \{ |a|^2 \frac{\partial}{\partial a} P[a^*,a] \bigg \} \nonumber \\ 
    &-& \frac{\hbar g^{K}}{2} \frac{\partial}{\partial a} \bigg \{ |a|^2 \frac{\partial}{\partial a^*} P[a^*,a] \bigg \}.
\end{eqnarray}
Furthermore, the fluctuation-dissipation theorem in the semi-classical approximation tells us that we have the relation $g^{K} = 2ig'' (k_B T/\hbar \bar{\omega})$. Using then a phase independent probability distribution $P[a^*,a] = P[|a|^2]$, we can easily see that both the terms with $-\hbar \delta$ as well as the terms with $\hbar g'$ (the real part of $\hbar g$) drop out. What is left is
\begin{eqnarray}
    0 &=& \bigg (\frac{k_B T}{\hbar \bar{\omega}} \bigg ) \bigg [ \frac{\partial}{\partial a^*} \bigg \{ |a|^2 \frac{\partial}{\partial a} P \bigg \} + \frac{\partial}{\partial a} \bigg \{ |a|^2 \frac{\partial}{\partial a^*} P \bigg \}
    \bigg ] \nonumber \\
    &+& 4 |a|^2 P + 2|a|^4 \frac{\partial P}{\partial |a|^2}.
\end{eqnarray}
Working this out further, we notice that 
\begin{dmath}
    \frac{\partial}{\partial a^*} \bigg \{ |a|^2 \frac{\partial}{\partial a} P \bigg \} + \frac{\partial}{\partial a} \bigg \{ |a|^2 \frac{\partial}{\partial a^*} P \bigg \}  \\
    = 4 |a|^2 \frac{\partial P}{\partial |a|^2} + 2|a|^4  \frac{\partial ^2 P}{(\partial |a|^2)^2},
\end{dmath}
which means that the stationary Fokker-Planck equation becomes
\begin{dmath}
0 = 2|a|^4\bigg \{ \bigg ( \frac{k_B T}{\hbar\bar{\omega}} \bigg ) \frac{\partial^2 P}{(\partial |a|^2)^2} + \frac{\partial P}{\partial |a|^2} \bigg \} + 4 |a|^2 \bigg \{ \bigg ( \frac{k_B T}{\hbar\bar{\omega}} \bigg ) \frac{\partial P}{\partial |a|^2} + P \bigg \}.
\end{dmath}
We therefore conclude that the probability distribution obeys simply
\begin{dmath}
\frac{\partial P}{\partial |a|^2} = - \frac{\hbar\bar{\omega}}{k_BT} P,
\end{dmath} 
which is the defining equation for the Gaussian ideal-gas solution
\begin{equation}
    P[a^*,a] \propto \exp \bigg ( - \frac{\hbar \bar{\omega}}{k_BT}  |a|^2 \bigg ),
\end{equation}
proportional to the expected classical Boltzmann factor. Note that there is no change in the phase-space volume in this case and there is no additional prefactor.

\section{Numerical simulation of semi-classical dynamics\label{num-sim}}
The Fokker-Planck equation of Eq.~(\ref{fpwhiteanddrive}) is given in terms of the eigenvalue $a$ of the annihilation operator and its complex conjugate $a^*$. The equation looks deceptively simple with the convection terms proportional to $\delta$, $g$, and $\omega_D$, and the diffusive term proportional to $g^K$. The probability distribution $P[a^*,a;t]$ can be represented on a grid and the spatial derivatives are evaluated using finite differences. The time-evolution of the probability distribution is then integrated forward in time using simple first-order finite difference in time. However, since $P$ is a real function, only terms on the right-hand side of Eq.~(\ref{fpwhiteanddrive}) that are imaginary contribute, whereas the real terms cancel. Furthermore, some of the terms are proportional to $|a|^2$ and since $a$ becomes large, these terms become large at the edges of the grid. At the same time the probability distribution can become narrow in some directions and thus the step size $\Delta a$ has to be chosen rather small. This makes the number of points in the grid large and precludes the use of implicit finite-difference methods. In the case of explicit finite-difference methods, the step size $\Delta t$ in time is bound by $\Delta a$ through the Courant condition, which requires small step sizes in time.

\newcommand{\prt}[2]{\frac{\partial #1}{\partial #2}}
\newcommand{\embr}[1]{\left( #1 \right)}
To avoid the cancellation of terms in the evolution of $P$, we use real coordinates $x=(a+a^*)/2$ and $y=(a-a^*)/2i$ and rewrite the Fokker-Planck equation in terms of these coordinates using
\begin{equation}
	\prt{}{a} =  \frac{1}{2} \left(\prt{}{x} - i \prt{}{y}\right) , \qquad \prt{}{a^*} =  \frac{1}{2} \left( \prt{}{x} + i \prt{}{y}\right).
\end{equation}
For the numerical integration, it is beneficial to write the evolution equation in the form of a conservation law and after some algabra we find
\begin{dmath}
\prt{}{t} P = - \prt{}{x} j_x^{P} - \prt{}{y} j_y^{P} ,
\end{dmath}
with the probability current densities
\begin{dmath}
j_x^{P} =  -\delta y P + \embr{g'y + g''x}\embr{x^2+y^2} P - \frac{\omega_D A_D}{4} y P +\frac{g^K}{4i}\embr{x^2+y^2}\prt{}{x} P , 
\end{dmath}
and
\begin{dmath}
j_y^{P} =  \delta x P + \embr{- g'x + g''y}\embr{x^2+y^2} P - \frac{\omega_D A_D}{4} x P +\frac{g^K}{4i}\embr{x^2+y^2}\prt{}{y} P,
\end{dmath}
where we remind ourselves that $g^K$ is purely imaginary. These equations are used for the simulations shown in Fig.~\ref{fig:trajectories}. 

There are many finite-difference methods proposed for this convection-diffusion problem. After some trial and error, we have selected the MacCormack method~\cite{MacCormack}, where the evolution is obtained after first making a predictor step, displacing the grid over half a grid spacing $\Delta a$ to the right, followed by a corrector step displacing the grid back by shifting it half a grid spacing to the left. This makes the method conditionally stable, compared to the simple forward time, centered space (FTCS) method. The stability is determined by the Courant-Friedrich-Lewy criterion, or Courant condition, given in our case by
\begin{equation}
	\frac{|j_x^{}|\Delta t}{P \Delta a} \leq 1, \qquad \frac{|j_y^{}|\Delta t}{P \Delta a} \leq 1. 
\end{equation}
Since the width of $P$ can become small in the evolution, we have to choose $\Delta a$ small and thus require a large number of time-steps for the evolution. 

The Fokker-Planck can also be rewritten in polar coordinates $\rho$ and $\phi$ and again after some algebra one arrives at
\begin{equation}
	\prt{}{t} P = -\frac{1}{\rho} \prt{}{\rho} \left( \rho j^{P}_\rho \right)  -\frac{1}{\rho} \prt{}{\phi}  j_\phi^{P} , \label{polar}
\end{equation}
with 
\begin{dmath}
	j_\rho^{P} = -\frac{\omega_D A_D}{4} \rho \sin 2 \phi P + g'' \rho^3 P + \frac{g^K}{4i} \rho^2 \prt{}{\rho} P , 
\end{dmath}
and
\begin{dmath}
	j_\phi^{P} =  \delta \rho P - \frac{\omega_D A_D}{4} \rho \cos 2 \phi P  - g' \rho^3 P + \frac{g^K}{4i} \rho \prt{}{\phi} P , 	
\end{dmath}
where we have expressed the right-hand side of Eq.~(\ref{polar}) in the form of a conservation law in polar coordinates. Note that the division by $\rho$ for each of the terms can be avoided by performing the integration in time using $\rho P(\rho,\phi)$ instead of $P(\rho,\phi)$. 

These equations allow us to find the location of the maximum of the equilibrium distribution using $j^P_\rho=0$ and $j^P_\phi=0$. The maximum is located at
\begin{equation}
	\langle \rho \rangle^2 = \frac{1}{|g|} \left( \delta \cos\phi_g + \sqrt{\left(\frac{\omega_D A_D}{4}\right)^2 - \delta^2 \sin^2 \phi_g} \right), 
\end{equation}
and
\begin{equation}
	\tan 2 \phi = \frac{g'' \langle \rho \rangle^2}{\delta - g' \langle \rho \rangle^2}. 
\end{equation}
The first result has already been derived by Smits {\it et al.}~\cite{exp}. In the case of small $g'$, these equations correspond to Eqs.~(\ref{evenwicht_phi}) and~(\ref{evenwicht_rho}) without the noise-induced drift that gives the small difference between the location of the maximum of the probability distribution and the average $\langle \rho \rangle$. 

%
%
%
%
%

\end{document}